\documentclass[a4paper,11pt]{article}
\pdfoutput=1 % if your are submitting a pdflatex (i.e. if you have
\usepackage{jheppub} % for details on the use of the package, please
\usepackage{fancyhdr}
\usepackage{graphicx}
\usepackage{xspace}
\usepackage{rotating}
\usepackage[normalem]{ulem}
\usepackage{braket}

\DeclareSymbolFont{extraup}{U}{zavm}{m}{n}
\DeclareMathSymbol{\varheart}{\mathalpha}{extraup}{86}
\DeclareMathSymbol{\vardiamond}{\mathalpha}{extraup}{87}

\usepackage{amsfonts}
\usepackage{pifont}

\newcommand{\be}{\begin{equation}}
\newcommand{\ee}{\end{equation}}
\newcommand{\bea}{\begin{eqnarray}}

\newcommand{\eea}{\end{eqnarray}}

\newcommand{\Rmnum}[1]{\expandafter\@slowromancap\romannumeral #1@}

\def\mev{\,{\rm MeV}\,}
\def\gev{\,{\rm GeV}\,}
\def\tev{\,{\rm TeV}\,}

\begin{document} 
  \begin{flushright}
    ADP--17--08/T1014
  \end{flushright}

%%%%%%%%%%%%%%%%%%%%%%%%%%%%%%%%%%%%%%%%%%%%
 \title{\boldmath %
Gravitational wave, collider and dark matter signals from a scalar singlet electroweak baryogenesis}
 %%%%%%%%%%%%%  Authors  Alphabetically %%%%%%%%%%%%%%%%%%
\author[a,1]{Ankit Beniwal,\note{ORCID ID: 0000-0003-4849-0611}}
\author[a,b,2]{Marek  Lewicki,\note{ORCID ID: 0000-0002-8378-0107}}
\author[c,d,3]{James D. Wells,\note{ORCID ID: 0000-0002-8943-5718}}
\author[a]{Martin White}
\author[a,4]{and Anthony G. Williams\note{ORCID ID: 0000-0002-1472-1592}}
%%%%%%%%%%%%  Affiliations %%%%%%%%%%%%%%%%%%%%%%%%%
\affiliation[a]{ARC Centre  of  Excellence  for Particle  Physics  at  the  Terascale
 (CoEPP) \& CSSM, Department of Physics, University of Adelaide, South Australia 5005, Adelaide, Australia}
\affiliation[b]{Faculty of Physics, University of Warsaw ul.\ Pasteura 5, 02-093 Warsaw, Poland}
\affiliation[c]{Michigan Center for Theoretical Physics, University of Michigan, Ann Arbor MI 48109, USA} 
\affiliation[d]{Deutsches Elektronen-Synchrotron DESY, Notkestra\ss e 85, D-22603 Hamburg, Germany}
%%%%%%%%%%%%%%%  Email Addresses %%%%%%%%%%%%%%%%%%%%
\emailAdd{ankit.beniwal@adelaide.edu.au}
\emailAdd{marek.lewicki@adelaide.edu.au}
\emailAdd{jwells@umich.edu}
\emailAdd{martin.white@adelaide.edu.au}
\emailAdd{anthony.williams@adelaide.edu.au}
%%%%%%%%%%%%%%%%%%%%%%%%%%%%%%%%%%%%%%%%%%%%
%  Abstract
\abstract{We analyse a simple extension of the SM with just an additional scalar singlet coupled to the Higgs boson. We discuss the possible probes for electroweak baryogenesis in this model including collider searches, gravitational wave and direct dark matter detection signals. We show that a large portion of the model parameter space exists where the observation of gravitational waves would allow detection while the indirect collider searches would not.}% and even the direct dark matter detection experiments would not.}


\arxivnumber{1702.06124}

\maketitle
\flushbottom

%%%%%%%%%%%%%%%%%%%%%%%%%%%%%%%%%%%%%%%%%%%%%%%%%%%%%%%%%%%%%%%%%%%%%%%

\section{Introduction} \label{sec:intro}

The discovery of the Higgs boson at the Large Hadron Collider (LHC)
\cite{Aad:2012tfa,Chatrchyan:2012xdj} has probably confirmed the existence of an elementary scalar and its role in the spontaneous breaking of the electroweak symmetry. It has also opened a new possibility for studying the details of the symmetry breaking from the properties of the Higgs boson. Another more recent observation of the first gravitational wave (GW) signal \cite{Abbott:2016blz} has given us a completely new way of probing the history of our universe. Specifically, rather violent events in the early history such as the electroweak phase transition should leave GW imprints. Yet another way of probing the early universe is via dark matter (DM). Current direct detection experiments are continually increasing their sensitivity to better probe the DM-nucleon scattering, and have placed strong exclusion limits on some of the allowed particle physics models with DM candidates. Motivated by these experimental probes that are continually developing, it is important to revisit the singlet scalar extension of the Standard Model (SM), a goal that we aim to achieve in this study.

%%%%%%%%%%%%%%%%%%%%%%%%%%%%%%%%%%%%%%%%%%%%%%%%%%%%%%%%%%%%%%%%%%%
We will focus on the two main features of this model. Firstly, it can facilitate electroweak baryogenesis (EWBG) \cite{Kuzmin:1985mm,Cohen:1993nk,Riotto:1999yt,Morrissey:2012db}, which aims to explain the observed baryon asymmetry of the universe through a strong first-order electroweak phase transition (EWPT). This phase transition is not first-order in the SM \cite{Arnold:1992rz,Kajantie:1996qd} and so a modification is needed to generate a barrier between the symmetric high temperature minimum and the electroweak symmetry breaking (EWSB) minimum as the universe cools down. A scalar singlet extension of the SM can provide this modification \cite{Curtin:2014jma,Kotwal:2016tex}, even though an effective theory with the new scalar integrated out suggests otherwise \cite{Damgaard:2015con,Damgaard:2013kva,Brauner:2016fla}. Secondly, after the $\mathbb{Z}_2$ symmetry is imposed, the new scalar serves as a viable DM candidate \cite{Silveira1985136,PhysRevD.50.3637,Burgess:2000yq}.

All of these attractive features are followed by equally attractive discovery prospects. Firstly, the new scalar inevitably modifies the Higgs potential which can be probed at collider experiments \cite{Curtin:2014jma,Katz:2014bha}. Secondly, a strong first-order phase transition generates a strong GW signal \cite{Grojean:2006bp}. This fact has been used in the literature to constrain various EWBG models \cite{Artymowski:2016tme,Vaskonen:2016yiu,Ashoorioon:2009nf,Enqvist:2014zqa,Tenkanen:2016idg,Huang:2016cjm,Kakizaki:2015wua,Hashino:2016rvx,Hashino:2016xoj,Kobakhidze:2016mch,Chala:2016ykx,Choi:1993cv,Dev:2016feu} and also in context of the zero temperature EW vacuum stability~\cite{Balazs:2016tbi,Khan:2014kba}. Lastly, the presence of a DM candidate with a non-zero abundance today provides strong direct detection limits on the model parameter space \cite{Cline:2013gha,Beniwal:2015sdl,He:2016mls}.  

The last possibility that we will explore comes from the fact that the early history of the universe is poorly constrained by astrophysical experiments. In order to identify all the parameter space in which the model can be viable, we will investigate how the allowed parameter space changes due to a modification of the cosmological history. The modification we consider comes simply from abandoning the assumption that the early universe was dominated by radiation. Instead, we will assume an additional energy constituent that redshifts faster than radiation. We will identify the experimental bounds on this scenario and show to what extent the modification changes the allowed parameter space of the scalar singlet model. 
This cosmological modification has two major effects. First on baryogenesis, as it helps to avoid the sphaleron bound \cite{Joyce:1996cp,Joyce:1997fc,Servant:2001jh,Lewicki:2016efe,Lewicki:2016oqx} and second on DM, as a faster expansion rate leads to an early freeze-out and consequently increases the resulting DM abundance today.   

The rest of the paper is organised as follows. In Section~\ref{sec:part_model}, we introduce the scalar singlet extension of the SM. The details of the EWBG in this model are given in Section~\ref{sec:baryo} along with the dynamics of the phase transition. In Section~\ref{sec:exp}, we discuss the discovery potential of the model, specially at colliders, gravitational wave and direct detection experiments. Section~\ref{sec:cosmomod} is devoted to the scenario of a modified cosmological history along with its impact on EWBG and DM abundance today. Our conclusions are presented in Section~\ref{sec:conclusions}. Details of the one-loop corrections to the effective potential are given in Appendix~\ref{sec:effpotappendix}.

%%%%%%%%%%%%%%%%%%%%%%%%%%%%%%%%%%%%%%%%%%%%%%%%%%%%%%%%%%%%
\section{Model}\label{sec:part_model}
One of the simplest extension of the SM is the addition of a new scalar singlet $S$ that couples to the SM Higgs boson. Assuming $\mathbb{Z}_2$ symmetry: $S \rightarrow -S$, the tree-level potential reads
\begin{equation}\label{eqn:pot0}
	V_{\textrm{tree}} (H,S) = -\mu^2|H|^2+ \lambda |H|^4+\lambda_{HS}|H|^2 S^2+\frac{1}{2} \mu_{S}^2 S^2+\frac{1}{4}\lambda_{S}S^4,
\end{equation}
where 
\begin{equation}
	H = \frac{1}{\sqrt{2}}
	\begin{pmatrix}
		\chi_1+i \chi_2 \\
		h + i \chi_3
	\end{pmatrix}
\end{equation}
and $\chi_{\{1,2,3\}}$ are the Goldstone bosons. Consequently, the potential in terms of $h$ and $S$ reads 
\begin{equation}\label{eqn:V0}
V_{\textrm{tree}}(h,S)=-\frac{1}{2}\mu^2 h^2 +\frac{1}{4}\lambda h^4+\frac{1}{2}\lambda_{HS} h^2 S^2+\frac{1}{2} \mu_{S}^2 S^2+\frac{1}{4}\lambda_{S}S^4.
\end{equation}
After electroweak symmetry breaking (EWSB), the physical mass of the new scalar $S$ is 
\begin{equation}\label{eqn:massh}
m^2_S =\mu_S^2 + \lambda_{HS}v_0^2,
\end{equation}
where $v_0 = \mu/\sqrt{\lambda}\approx 246$\gev is the SM Higgs VEV. At tree-level, the Higgs mass and its VEV fixes the constants $\mu$ and $\lambda$ in Eq.~\eqref{eqn:V0} to $m_h=\sqrt{2}\mu=125$\gev and $\lambda=m_h^2/2v_0^2\approx 0.129$ respectively. We adopt renormalisation conditions that do not modify these values.

We will discuss a wide range of the scalar-Higgs coupling $\lambda_{HS}$\footnote{ We show our results for $\lambda_{HS}\in [0.2,4\pi]$, however, one has to remember that for values of the coupling larger than a few~\cite{Curtin:2014jma}, the one-loop corrections become unreliable.} for the scalar masses $m_S$ above $m_h/2$.\footnote{In the region $m_S < m_h/2$, values of $\lambda_{HS}$ that are of interest for EWBG are mostly excluded by the limits on the Higgs invisible branching ratio (see e.g., Ref.~\cite{Beniwal:2015sdl}).} In all of our plots, we fix the scalar self-coupling $\lambda_S=1$. The dependence of most of our results on $\lambda_S$ is rather mild; an increase in its value would only shift the allowed region to slightly higher values of $\lambda_{HS}$. We include one-loop corrections to the potential at zero and finite temperature (see Appendix~\ref{sec:effpotappendix} for more details). The most important effect of these corrections is the appearance of a barrier between the symmetric phase at $\langle h \rangle = 0$ and the EWSB one at $\langle h \rangle >0$. 

\section{Electroweak baryogenesis}\label{sec:baryo}
In the early universe and at very high temperatures, thermal corrections to the scalar potential restore the electroweak symmetry. As the universe cools down, the EWSB minimum emerges. Due to the corrections from the new scalar, the electroweak minimum can be separated from the symmetric one at $\langle h \rangle =0$ by a potential barrier, thereby allowing a first-order phase transition that is absent in the SM \cite{Kajantie:1996qd}.

The necessary condition  for the EWBG that we will focus on is the decoupling of the sphaleron processes after the EWPT. The sphaleron processes present in the SM are connected with $SU(2)$ gauge interactions and provide baryon number $(B)$ violation necessary to create baryon-anti-baryon asymmetry. However, if they are not decoupled after the transition, they quickly wash-out any previously created asymmetry. As $SU(2)$ interactions, they are heavily suppressed once the electroweak symmetry is broken. This breaking is quantified by the Higgs VEV. Thus, the decoupling of the sphalerons leads to the following well-known condition
\begin{equation}\label{eqn:sphbound}
	\frac{v}{T}\geq 1,
\end{equation}
where $v$ is the Higgs VEV calculated at temperature $T$. We will start with a generic discussion in subsection~\ref{sec:vacuumstructure} approximating the transition temperature as the critical temperature $T_c$ at which the minima of the potential are degenerate. In the next subsection~\ref{sec:dynamicsofthephasetransition} we will discuss the dynamics of the transition and calculation of the temperature $T_*$ at which the transition truly begins. The calculation of the sphaleron rate is technically complicated and therefore leads to slightly different bounds on $v/T$, as present in Refs.~\cite{Katz:2014bha,Quiros:1999jp,Funakubo:2009eg,Fuyuto:2014yia}. For simplicity, we will simply employ the above bound. 

\subsection{Vacuum structure}\label{sec:vacuumstructure}
%%%%%%%%%%%%%%%%%%%%%%%%%%%%%%%%%%%%%%%%%%%%%%%%%%%%%%%%%%%%%%%%%%%%%%
\begin{figure}[t]
	\includegraphics[width=\textwidth]{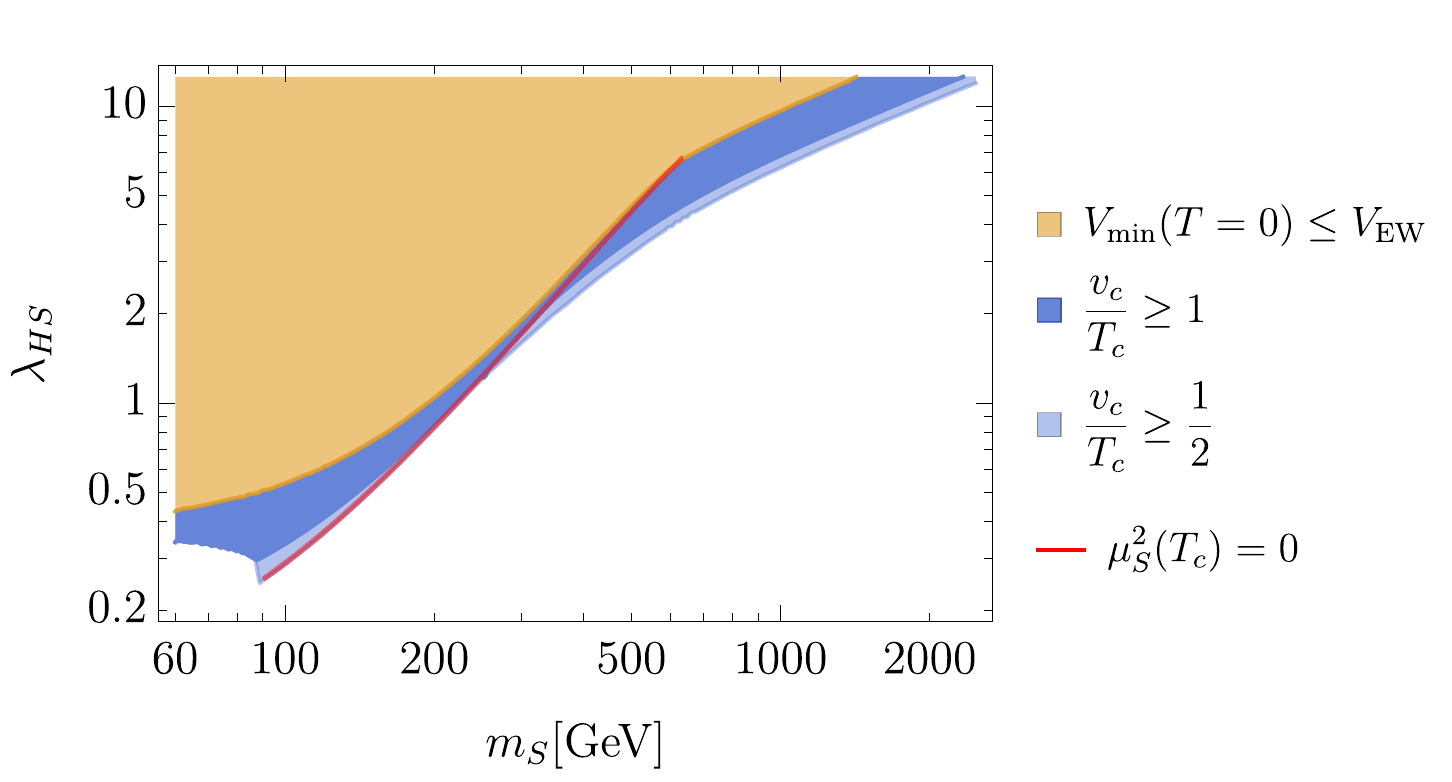}
	\caption{Parameter space of the scalar singlet model relevant for EWBG. The yellow region is excluded because in that region, the electroweak minimum is not the global minimum at zero temperature. The blue region realises a strong first-order phase transition whereas the light blue region can still be allowed due to the cosmological modification. The solid red line marks the boundary between the regions where $\mu_S^2 (T_c) < 0$ and $\mu_S^2 (T_c) > 0$ (see text for more details).}
	\label{fig:EWBGplot}
\end{figure}
%%%%%%%%%%%%%%%%%%%%%%%%%%%%%%%%%%%%%%%%%%%%%%%%%%%%%%%%%%%%%%%%%%%%%%
Before going into the details of the transition dynamics, we will discuss the parameter space allowed by the vacuum structure. Fig.~\ref{fig:EWBGplot} shows the relevant regions of the model parameter space where the minimum at the origin and the EWSB minimum are separated by a barrier at the critical temperature $T_c$. The yellow region is excluded because in this region, the electroweak minimum is \emph{not} the true minimum of the potential at zero temperature. For low masses, this happens because the minimum in the $S$  direction is deeper, whereas for very large couplings, the electroweak minimum is pushed up by quantum corrections to values above the minimum at the origin.  In both cases the universe will never transition to the broken EW symmetry phase and thus these situations are excluded. The region of very small mass and coupling is also excluded because the negative mass terms start to overpower the $h^2S^2$ coupling and the new minimum appears between the minima in the $h$ and $S$ directions at $\langle S \rangle>0, \langle h \rangle>0$. 
  
Depending on the sign of $\mu_S^2$ in this model, the EWPT can proceed in two ways.
\begin{enumerate}
	\item $\mu_S^2 > 0$: This occurs at large $m_S$ and small $\lambda_{HS}$. In this case, the potential grows as we move away from $S = 0$. We can thus discuss only the one-dimensional potential along the $h$ direction. This leads to a one-step phase transition during which the field is initially in a homogeneous configuration at the origin and tunnels through the barrier towards the electroweak minimum.
	\item $\mu_S^2 < 0$: This occurs at small $m_S$ and large $\lambda_{HS}$. In this case, the universe can transition into a minimum along the $S$ direction before EWPT occurs. As discussed below, this scenario requires a precise numerical calculation to calculate the exact details of the EWPT.
\end{enumerate}

%%%%%%%%%%%%%%%%%%%%%%%%%%%%%%%%%%%%%%%%%%%%%%%%%%%%%%%%%%%%%%%%%%%%%
%%%%%%%%%%%%%%%%%%%%%%%%%%%%%%%%%%%%%%%%%%%%%%%%%%%%%%%%%%%%%%%%%%%%%%
%%%%%%%%%%%%%%%%%%%%%%%%%%%%%%%%%%%%%%%%%%%%%%%%%%%%%%%%%%%%%%%%%%%%%%
\subsection{Dynamics of the phase transition}\label{sec:dynamicsofthephasetransition}
%%%%%%%%%%%%%%%%%%%%%%%%%%%%%%%%%%%%%%%%%%%%%%%%%%%%%%%%%%%%%%%%%%%%%%
%%%%%%%%%%%%%%%%%%%%%%%%%%%%%%%%%%%%%%%%%%%%%%%%%%%%%%%%%%%%%%%%%%%%%%
The EWPT occurs after the temperature of the universe drops below the critical temperature and the minimum with non-zero Higgs VEV becomes the global minimum. We will discuss a first-order phase transition during which this new global minimum is separated from the electroweak symmetry preserving minimum by a potential barrier.

We now explain the dynamics of the phase transition in more detail. In the early universe, the EW phase transition is driven by thermal fluctuations that eventually excite the field enough to cross the potential barrier. Calculating the details of the phase transition essentially boils down to finding the field profile corresponding to such thermal excitations that will appear most quickly and drive the transition. The crucial quantity for finding the temperature at which the transition proceeds is the probability of finding a field configuration with action $S_3$ within a volume $\mathcal{V}$ \cite{Linde:1981zj, Linde:1980tt}
\begin{equation}\label{eq:decaywidth}
\frac{\Gamma}{\mathcal{V}} \approx T^4 \exp\left(-\frac{S_3(T)}{T}\right).
\end{equation}
Thus, the most probable configurations (as usual) are those with the smallest action, which in turn are the most symmetric ones. Noticing also that we can start with a static field configuration (as the time derivative could only increase the result), we can write down the action of our $O(3)$ symmetric field bubble. We will also be interested in cases where the $S$ field cannot be neglected during the transition leading to a slightly more complicated action involving both fields as 
\begin{equation}\label{eq:actionfunc}
S_3=4\pi \int dr\, r^2\left\{\frac{1}{2}\left(\frac{d h}{dr}\right)^2+\frac{1}{2}\left(\frac{d S}{dr}\right)^2+V_{\textrm{eff}}(h,S,T)\right\},
\end{equation}
where $V_{\textrm{eff}}$ is the effective potential (see Appendix \ref{sec:effpotappendix} for more details).

In the simple case where the $m_S$ is large, the potential grows quickly in the $S$ direction and we can set $S=0$, leading to a much simpler analysis involving only the Higgs field direction \cite{Lewicki:2016efe,Artymowski:2016tme}. However for small $m_S$, the universe transitions to the $\langle S\rangle > 0$, $\langle h \rangle=0$ minimum before the EWPT. During this first transition, no barrier is generated between the origin and the $\langle S \rangle>0$ vacuum and so it is a smooth crossover. The new problem when compared with the single field case is finding a trajectory in the field space that connects the initial vacuum $(\langle S\rangle>0, \langle h \rangle=0)$ with the electroweak one $(\langle S\rangle=0$, $\langle h \rangle=v_0)$ and minimises the action in Eq.~\eqref{eq:actionfunc}.

We follow an approach similar to the one outlined in Refs.~\cite{Cline:1999wi,Profumo:2010kp,Wainwright:2011kj}. We begin by choosing a path $\vec{\phi}(t)=\left(h(t),S(t)\right)$ that connects the initial and final vacuum. We always set 
\begin{equation}
	\left|\frac{d \vec{\phi}}{d t}\right|^2=\left(\frac{d h}{d t}\right)^2+\left(\frac{d S}{d t}\right)^2=1
\end{equation}
such that $d \vec{\phi}/dt$ is a unit vector parallel to the path whereas $d^2 \vec{\phi}/dt^2$ is perpendicular to the path. We can rewrite the equations of motion (EOMs) from the original action in Eq.~\eqref{eq:actionfunc}
\begin{equation}\label{eq:eom0}
\frac{d^2 h}{d r^2}+\frac{2}{r}\frac{d h}{d r}=\frac{\partial V}{\partial h}, \quad
\frac{d^2 S}{d r^2}+\frac{2}{r}\frac{d S}{d r}=\frac{\partial V}{\partial S},
\end{equation}
in terms of the path $\vec{\phi}(t)$ as
\begin{equation}\label{eq:eompath}
\frac{d \vec{\phi}}{d t}\frac{d^2 t}{d r^2}+\frac{d^2 \vec{\phi}}{d t^2}\left(\frac{d t}{d r}\right)^2+\frac{2}{r}\frac{d \vec{\phi}}{d t}\frac{d t}{d r}=\nabla V.
\end{equation}
Now, taking the part proportional to $d \vec{\phi}/dt$ gives us the EOM along the path
\begin{equation}\label{eq:eomparallel}
\frac{d \vec{\phi}}{d t}\left(\frac{d^2 t}{d r^2}+\frac{2}{r}\frac{d t}{d r}\right)=\left( \nabla V\right)_{\parallel}, 
\end{equation}
whereas taking the part proportional to $d^2 \vec{\phi}/dt^2$ gives the EOM perpendicular to the path
\begin{equation}\label{eq:eomperp}
\frac{d^2 \vec{\phi}}{d t^2}\left(\frac{d t}{d r}\right)^2=\left(\nabla V\right)_{\perp}.
\end{equation}
For a given path (just as in the one-dimensional case), finding the bubble profile means solving  Eq.~\eqref{eq:eomparallel} along the path
\begin{equation}\label{eq:eombubble}
 \frac{d^2 t}{d r^2}+\frac{2}{r}\frac{d t}{d r}=\frac{d V}{d t}
\end{equation}
to find $t(r)$ satisfying the following boundary conditions needed for a finite action
\begin{equation}\label{eq:bubbleboundary}
\left. \frac{d t}{d r}\right|_{r=0}=0, \quad t(r\rightarrow\infty)= V_f,
\end{equation} 
where $V_f$ is the value of the potential at the decaying initial vacuum. The problem in choosing a certain path is that we completely neglect Eq.~\eqref{eq:eomperp} which should also be satisfied if one wants to find a solution of Eq.~\eqref{eq:eom0}.

Our approach to solve both EOMs is the following. We choose a certain initial path and solve Eq.~\eqref{eq:eombubble} to satisfy the boundary conditions in Eq.~\eqref{eq:bubbleboundary}. This gives us $dt/dr$ along the path and allows us to calculate
\begin{equation}
\vec{N}=\frac{d^2 \vec{\phi}}{d t^2}\left(\frac{d t}{d r}\right)^2-\left(\nabla V\right)_{\perp}.
\end{equation}
Now, we modify our path to obtain $\vec{N}=0$, which corresponds to finding a solution of Eq.~\eqref{eq:eomperp}. In practice, we have to do this iteratively. Each step consists of moving each point along our path in the direction of $\vec{N}$ and finding a modified path by fitting a polynomial to the modified points. Fitting a function is necessary as otherwise this algorithm becomes highly unstable. This is because the result of one such modification is not a smooth function and the second derivative can grow uncontrollably, which would lead to an even bigger growth in subsequent modifications. We choose to fit a polynomial of order $5$, and have checked that using higher powers does not increase the accuracy of the result any further. After 20 such modifications, we again calculate the tunnelling action along the modified path by solving Eq.~\eqref{eq:eombubble}. This gives us the next approximation of the $S_3$ and $dt/dr$ along the path for further path modification. After a few such steps, the action stabilises which means a solution has been found.

We have checked that the above algorithm converges to the same result with any reasonable initial guess for the path. However, in practice it is most convenient to start with a path that is obtained by choosing $S$ that minimises the potential for each $h$ between the initial and final vacuum. In fact, in this model, this simple choice proves to be a very good approximation and the path obtained using the path modification algorithm decreases the resulting action only by a few percent. This leads to a negligible modification of the transition temperature $T_*$.

%%%%%%%%%%%%%%%%%%%%%%%%%%%%%%%%%%%%%%%%%%%%%%%%%%%%%%%%%%%%%%%%%%%%%
\begin{figure}[t]
	\includegraphics[width=\textwidth]{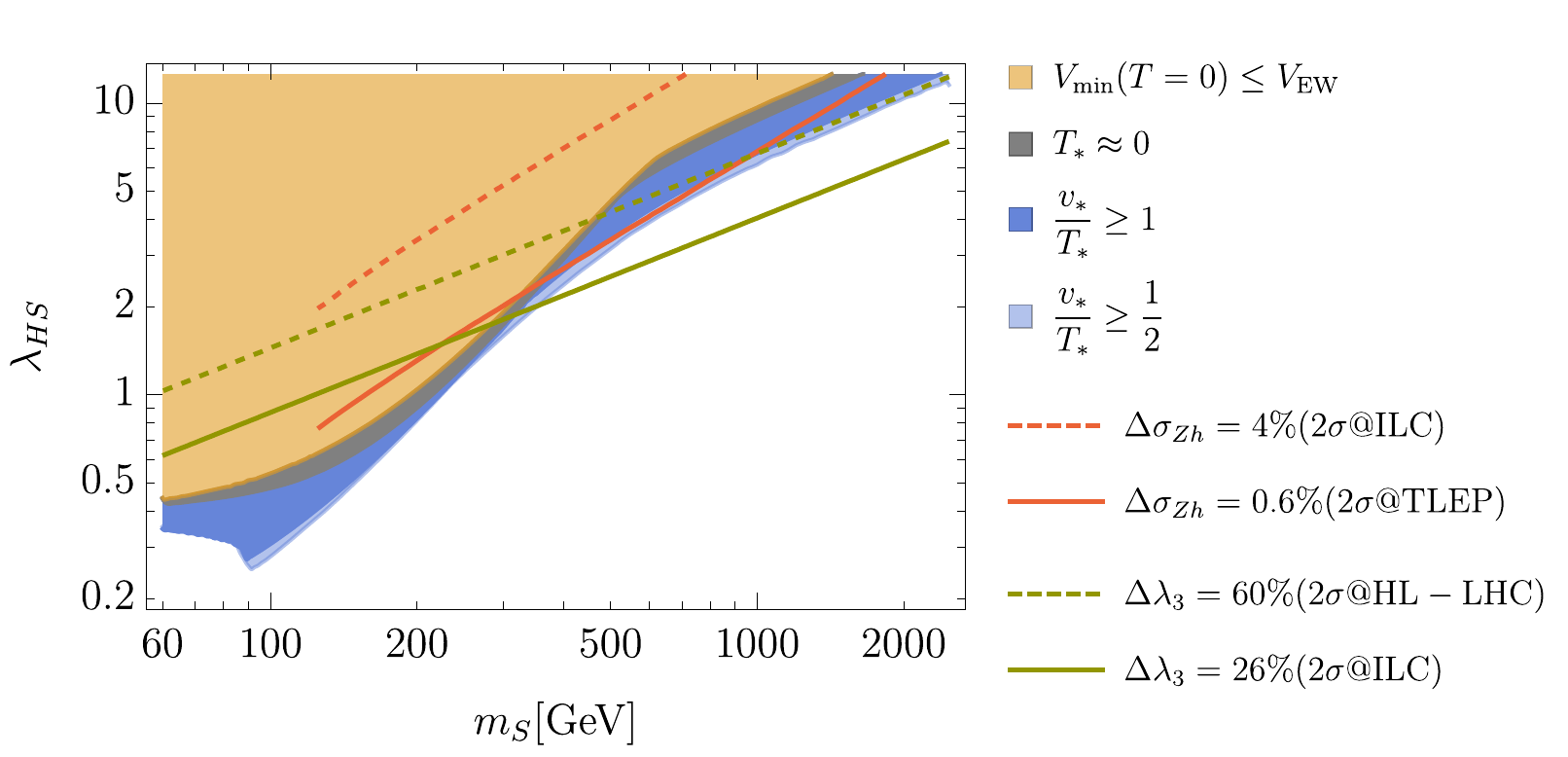}
	\caption{Parameter space of the scalar singlet model relevant for EWBG along with the reach of various collider experiments. The yellow shaded region is excluded because in that region, the electroweak minimum is not the global minimum at zero temperature. In the grey region, the universe is trapped in a metastable vacuum that preserves electroweak symmetry. The blue region realises a strong first-order phase transition whereas the light blue region can still be allowed due to the cosmological modification. 	
	Regions above the dotted and dashed lines will be accessible at colliders. Here $	\Delta \lambda_3 \equiv (\lambda^{\textrm{SM}}_{3} - \lambda_3)/\lambda^{\textrm{SM}}_{3}$ is the modification of the triple Higgs coupling with respect to the SM.}
	\label{fig:EWBGexclplot2}
\end{figure}
%%%%%%%%%%%%%%%%%%%%%%%%%%%%%%%%%%%%%%%%%%%%%%%%%%%%%%%%%%%%%%%%%%%%%

Now, we are ready to use the action in Eq.~\eqref{eq:actionfunc} and the decay width in Eq.~\eqref{eq:decaywidth} to find $T_*$. We assume that the phase transition proceeds when at least one bubble is nucleated in every horizon, i.e.,
\begin{equation}\label{eqn:TnucleationRAD} 
%\int_{0}^{t_*}\Gamma_{\rm tot}\, dt  =
\int_{T_*}^{\infty} \frac{dT}{T}\frac{1}{H} \Gamma V_H =\int_{T_*}^{\infty}\frac{dT}{T} \left(\frac{1}{2\pi}\sqrt{\frac{45}{\pi g_{\rm eff}}} \frac{M_p}{T} \right)^4 \exp\left(-\frac{S_3(T)}{T}\right)=1,
\end{equation}
where $H$ is the Hubble rate, $V_H$ is the horizon volume and $g_{\rm eff}$ is the effective number of degrees of freedom at temperature $T$.\footnote{We use the tabulated values of $g_{\rm eff}$ as a function of $T$ from \texttt{micrOMEGAs\_3.6.9.2} \cite{Belanger:2014vza}.} Under this assumption, our result depends on the thermal history of the universe. Indeed, this dependence is not negligibly small, as previously shown in Ref.~\cite{Lewicki:2016efe}.

In Figs.~\ref{fig:EWBGplot} and \ref{fig:EWBGexclplot2}, we show the model parameter space relevant for EWBG. The main difference between the simplified analysis using the critical temperature $T_c$ and the actual transition temperature $T_*$ is visible in the very strong transition region (i.e., $v/T \geq 1$). Large values of $v_c$ mean that the barrier between the electroweak minimum and the symmetric one is very wide; the probability of the transition is so low that the universe would remain until today (i.e., $T_* \sim 0$) in the vacuum that preserves EW symmetry, which is of course excluded.  

%%%%%%%%%%%%%%%%%%%%%%%%%%%%%%%%%%%%%%%%%%%%%%%%%%%%%%%%%%%%%%%%%%%%%
%%%%%%%%%%%%%%%%%%%%%%%%%%%%%%%%%%%%%%%%%%%%%%%%%%%%%%%%%%%%%%%%%%%%%

%\begin{figure}[t]
%	%\includegraphics[height=5.4cm]{EWBGplot2.pdf} 
%	\includegraphics[width=\textwidth]{EWBGlogplot2.pdf}
%	\caption{Parameter space of the scalar singlet model relevant for baryogenesis. The yellow shaded region is excluded because the electroweak minimum is not the global minimum at zero temperature. In the grey region, the universe is trapped in the metastable vacuum which preserves electroweak symmetry. The blue region realises a strong first-order phase transition and the light blue region can still be allowed due to the cosmological modification.} 
%	\label{fig:EWBGplot2}
%\end{figure}

%%%%%%%%%%%%%%%%%%%%%%%%%%%%%%%%%%%%%%%%%%%%%%%%%%%%%%%%%%%%%%%%%%%%%%
%\begin{figure}[t]
%	\centering
%	%\includegraphics[height=4.4cm]{EWBGexclplot2.pdf} 
%	\includegraphics[width=\textwidth]{EWBGlogexclplot2.pdf}
%	\caption{Parameter space of the scalar singlet model capable of realising EWBG together with the reach of various collider experiments.} 
%	\label{fig:EWBGexclplot2}
%\end{figure}
%%%%%%%%%%%%%%%%%%%%%%%%%%%%%%%%%%%%%%%%%%%%%%%%%%%%%%%%%%%%%%%%%%%%%

%%%%%%%%%%%%%%%%%%%%%%%%%%%%%%%%%%%%%%%%%%%%%%%%%%%%%%%%%%%%%%%%%%%%%
%%%%%%%%%%%%%%%%%%%%%%%%%%%%%%%%%%%%%%%%%%%%%%%%%%%%%%%%%%%%%%%%%%%%%
\section{Experimental probes}\label{sec:exp}
In this section, we discuss the various experimental probes for scalar singlet EWBG.

\subsection{Collider signals}\label{sec:collidersignals}
Although a direct detection of the new scalar at the LHC is hopeless due to the small signal to background ratio, it could be possible at a $100$\tev collider, provided its mass (coupling) is small (large) enough \cite{Curtin:2014jma}. This distinct possibility, however, covers only a small portion of the parameter space that is of interest to us; indirect collider searches prove a far better probe of this scenario.

The first indirect probe comes from the modification of the triple Higgs coupling. This modification comes from the new scalar $S$ and can be easily obtained by differentiating the potential including one-loop contribution from $S$ in Eq.~\eqref{eqn:V1}. This gives the following result
\begin{equation}
\lambda_3= \left. \frac{1}{6}\frac{\partial^3 V(h,S=0,T=0)}{\partial h^3} \right|_{h=v_0} \approx \frac{m_h^2}{2 v_0}+ \frac{\lambda_{H S}^3 v_0^3}{24 \pi^2 m_S^2}.
\end{equation}
This coupling can only be measured at the HL-LHC in double Higgs production events where the very low cross section again makes the measurement difficult. The estimated precision on this coupling is about $30\%$ at the HL-LHC \cite{Goertz:2013kp},
% but can get to $10\%$ at a $100$ TeV $pp$ collider with $30$\,ab$^{-1}$ of data \cite{Barr:2014sga}.
% Similar precision can be achieved at lepton colliders, for instance, at 1\,TeV ILC with $2.5$\,ab$^{-1}$ where the accuracy is predicted to be $13\%$ \cite{Asner:2013psa}. 
but can get to $13\%$ at 1\,TeV ILC with $2.5$\,ab$^{-1}$ \cite{Asner:2013psa}. 
Much better precision could be reached at the $100$ TeV $pp$ collider~\cite{Contino:2016spe,Barr:2014sga}. Together with direct detection of the new scalar produced through an off-shell Higgs \cite{Curtin:2014jma} could probe the whole relevant parameter space. However, the $100$ TeV collider has a much bigger time frame than other discussed experiments. Therefore, it will not be included in our comparisons.

The second possibility for indirect detection of the new scalar $S$ is through its modification to the $Z h$ production at lepton colliders. The fractional change relative to its SM value is given by \cite{Englert:2013tya,Curtin:2014jma}
 \begin{equation}
\Delta \sigma_{Zh}=\frac{1}{2}\frac{\lambda_{HS}^2 v_0^2}{4\pi^2 m_h^2}\left[ 1 + F\left(\frac{m_h^2}{4m_S^2}\right)\right],
\end{equation}
where
\begin{equation}
F(\tau)=\frac{1}{4\sqrt{\tau(\tau-1)}}\log \left( \frac{1-2\tau-2\sqrt{\tau(\tau-1)}}{1-2\tau+2\sqrt{\tau(\tau-1)}} \right).
\end{equation}
ILC can achieve a precision of $2\%$ whereas FCC-ee/TLEP will be able probe it with $0.6\%$ accuracy at the $95\%$ C.L. \cite{Dawson:2013bba}.

In Fig.~\ref{fig:EWBGexclplot2}, we show parts of the model parameter space that are accessible at colliders. Clearly, a measurement of $\lambda_3$ is the best probe of the neutral scalar scenario. The ILC and a $100$\tev $pp$ collider would be able to probe most of the strong first-order PT model parameter space for scalar masses above $\sim 350$\gev. The $Zh$ production is a somewhat weaker probe. The ILC should not see any modification if our model is realised since it can only probe the unphysical parameter space. FCC-ee/TLEP on the other hand could probe a significant part of the parameter space where a one-step phase transition can occur. However, it still has a smaller reach in the low mass region than the ILC. In the very high mass region, it cannot probe the full parameter range where a strong phase transition occurs.

%%%%%%%%%%%%%%%%%%%%%%%%%%%%%%%%%%%%%%%%%%%%%%%%%%%%%%%%%%%%%%%%%%%%%%
%%%%%%%%%%%%%%%%%%%%%%%%%%%%%%%%%%%%%%%%%%%%%%%%%%%%%%%%%%%%%%%%%%%%%%

\subsection{Gravitational wave signals}\label{sec:GWs}
A first-order phase transition is a very violent event in the history of the universe. Nucleation and subsequent collisions of bubbles converting the symmetric vacuum to the electroweak one is a process that is very far from equilibrium and brings about large transfers of energy. Seeing as all the fields are flat and interact gravitationally, it is the perfect setting for the creation of gravitational waves. This issue has been widely discussed in the literature where three main sources of GWs have been identified.
These are the collisions of the bubble walls \cite{Kamionkowski:1993fg,Huber:2008hg,Jinno:2016vai}, sound waves generated after the transition \cite{Hindmarsh:2013xza,Hindmarsh:2015qta}, and the magneto-hydrodynamical (MHD) turbulence in the plasma \cite{Caprini:2009yp}.

The details of the phase transition described in the previous section allow us to calculate the energy carried by the bubbles which drive the transition and the time scale in which it will proceed. These quantities are exactly what we need to obtain the GW signals produced by the transition \cite{Grojean:2006bp}. The first parameter crucial for the GW spectrum is the ratio of released latent heat from the transition to the energy density of the plasma background \cite{Caprini:2015zlo}
\begin{equation}\label{eqn:alpha}
\alpha= \left.\frac{1}{\rho_{R}}\left[-(V_{\rm EW}-V_f)+ T \left(\frac{dV_{\rm EW}}{dT} - \frac{dV_f}{dT}\right)\right]\right|_{T=T_*} ,
\end{equation}
where $V_f$ is the value of the potential in the unstable vacuum (in which the field initially resides) and $V_{\rm EW}$ is the value of the potential in the final vacuum (in which the electroweak symmetry is broken). The inverse time of the phase transition is given by
\begin{equation}\label{eqn:beta}
\frac{\beta}{H}= \left. \left[T \frac{d}{dT} \left(\frac{S_3(T)}{T} \right)\right]\right|_{T=T_*}.
\end{equation}
The parameters $\alpha$ and $\beta$ in Eqs.~\eqref{eqn:alpha} and \eqref{eqn:beta} respectively allows us to calculate the GW signals produced during the phase transition.

The first important source of GWs is bubble collisions. Peak frequency of the resulting signal is  \cite{Huber:2008hg}
\begin{equation}
f_{\textrm{col}}=
16.5\times 10^{-6} \frac{0.62}{v_b^2-0.1 v_b+1.8}\frac{\beta}{H}
\frac{T_*}{100} \left(\frac{g_*}{100}\right)^{\frac{1}{6}} {\rm Hz}
\end{equation}
with the following energy density
\begin{equation}
\Omega h^2_{\textrm{col}}(f)=1.67\times 10^{-5}\left(\frac{\beta}{H}\right)^{-2}
\frac{0.11 v_b^3}{0.42+v_b^2}
\left(\frac{\kappa \alpha }{1+\alpha }\right)^2 
\left(\frac{g_*}{100}\right)^{-\frac{1}{3}}
\frac{3.8 \left(f/f_{\rm col}\right)^{2.8}}{1+2.8 \left(f/f_{\rm col}\right)^{3.8}},
\end{equation}
where the efficiency factor $\kappa$ and the bubble wall velocity $v_b$ are given by
\begin{equation}\label{eq:bubblespeed}
v_b=\frac{1/ \sqrt{3}+\sqrt{\alpha ^2+2 \alpha /3}}{1+\alpha }, \quad
%\kappa=\frac{1}{1+0.715 \alpha }\left(0.715 \alpha +\frac{4}{27} \sqrt{\frac{3 \alpha }{2}}\right).
\kappa=\frac{\alpha_{\infty}}{\alpha}\left( \frac{\alpha_{\infty}}{0.73+0.083\sqrt{\alpha_{\infty}}+\alpha_{\infty}} \right).
\end{equation}
%%%%%%%%%%%%%%%%%%%%%%%%%%%%%%%%%%%%%%%%%%%%%%%%%%%%%%%%%%%%
\begin{figure}[t] 
	\centering
	\includegraphics[width=\textwidth]{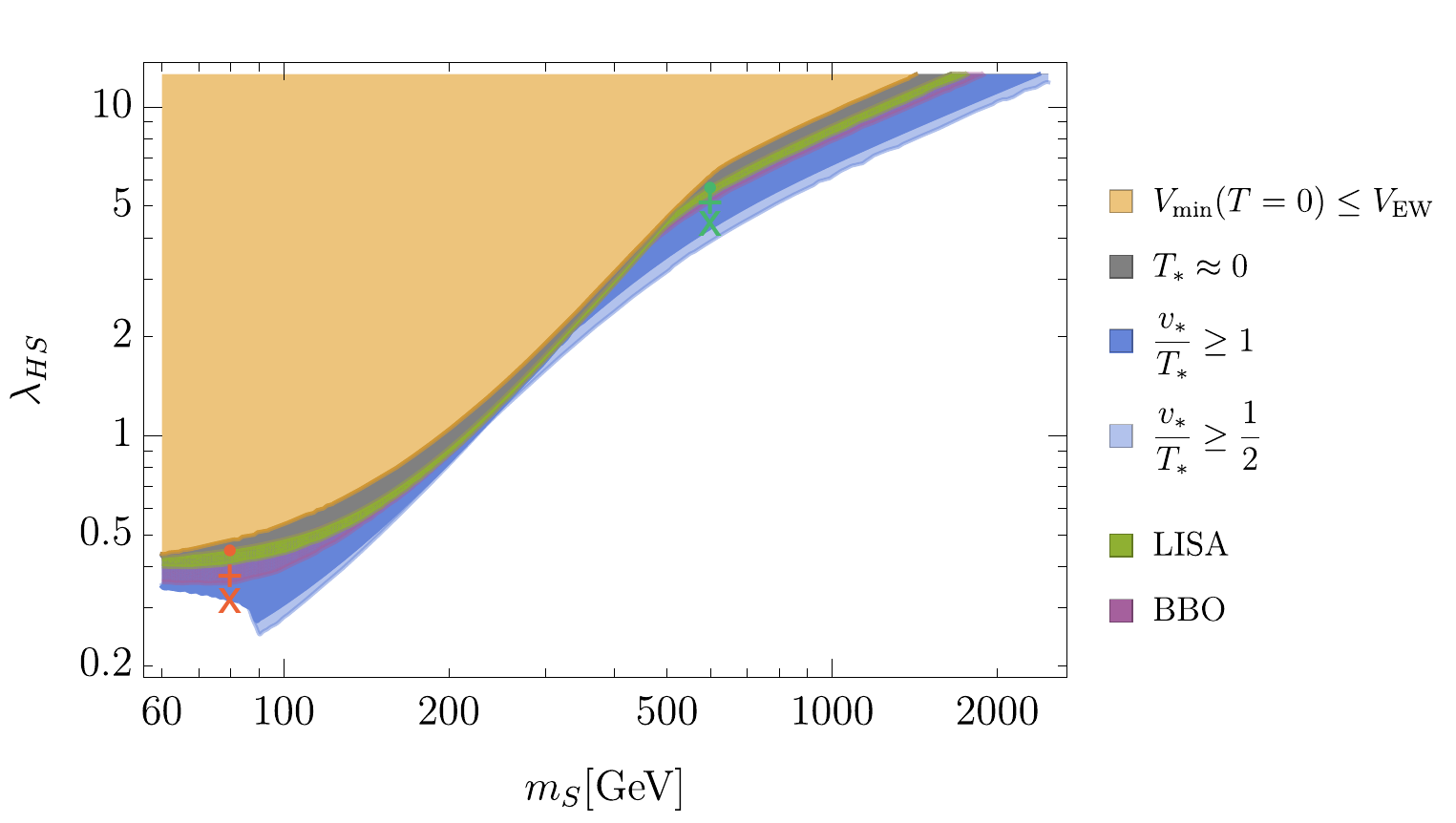} 
	\caption{Parameter space of the scalar singlet model relevant for EWBG.
In the green and purple regions GW signals produced during the phase transition will be accessible in future experiments such as LISA and BBO respectively.
	 A few example points are also highlighted to match with their GW spectra in Fig.~\ref{fig:GWplot}.}
	\label{fig:GWdetectionplot}
\end{figure}
%%%%%%%%%%%%%%%%%%%%%%%%%%%%%%%%%%%%%%%%%%%%%%%%%%%%%%%%%%%%

This definition already includes the fact that during a very strong phase transition, the energy deposited into the fluid saturates at \cite{Espinosa:2010hh,Bodeker:2009qy,Megevand:2009gh}
\begin{equation}
\alpha_{\infty}=0.49 \times 10^{-3} \left(\frac{v_*}{T_*}\right)^2.
\end{equation}
We obtain values of $\alpha \in [10^{-3}, 10]$ and $\beta/H \in [1, 10^4]$. The condition $\alpha > \alpha_\infty$ is satisfied by the majority of points giving hope for detection in near future. The bubble wall velocity in Eq.~\eqref{eq:bubblespeed} provides only a lower bound on the true wall velocity \cite{Huber:2008hg}. However, we have checked that replacing this with $v_b=1$, which is more appropriate for a very strong transition, does not modify our results noticeably. The same can be said with respect to varying the bubble wall velocity within some uncertainty, say $20\%$. Although the specific spectra change slightly as their frequency and magnitude is multiplied by this $\mathcal{O}(1)$ factor, the resulting reach of future GW experiments does not change significantly. Also, for points satisfying $\alpha<\alpha_{\infty}$, the contribution of bubble collisions to the GW signal can be neglected.

The second important source of GWs are sound waves created in the plasma after the bubbles collide. The corresponding peak frequency is \cite{Hindmarsh:2013xza,Hindmarsh:2015qta}
\begin{equation}
f_{\rm sw}=1.9 \times 10^{-5} \frac{\beta}{H} \frac{1}{v_b} \frac{T_*}{100}\left({\frac{g_*}{100}}\right)^{\frac{1}{6}} {\rm Hz }
\end{equation}
with the following energy density
\begin{equation}
\Omega h^2_{\rm sw}(f)=2.65 \times 10^{-6}\left(\frac{\beta}{H}\right)^{-1}
\left(\frac{\kappa \alpha }{1+\alpha }\right)^2 
\left(\frac{g_*}{100}\right)^{-\frac{1}{3}}
v_b
\left(\frac{f}{f_{\rm sw}}\right)^3 \left(\frac{7}{4+3 \left(f/f_{\rm sw}\right)^2}\right)^{7/2}.
\end{equation}

The last important source of GW signals is MHD turbulence in the plasma. The frequency at the peak of this contribution is \cite{Caprini:2009yp}
\begin{equation}f_{\rm turb}=2.7  \times 10^{-5}
\frac{\beta}{H} \frac{1}{v_b} \frac{T_*}{100}\left({\frac{g_*}{100}}\right)^{\frac{1}{6}} {\rm Hz }
\end{equation} 
with the following energy density
\begin{equation}
\Omega h^2_{\rm turb}(f)=3.35 \times 10^{-4}\left(\frac{\beta}{H}\right)^{-1}
\left(\frac{\epsilon \kappa \alpha }{1+\alpha }\right)^{\frac{3}{2}} 
\left(\frac{g_*}{100}\right)^{-\frac{1}{3}}
v_b
\frac{\left(f/f_{\rm turb}\right)^3\left(1+f/f_{\rm turb}\right)^{-\frac{11}{3}}}{\left[1+8\pi f a_0/(a_* H_*)\right]},
\end{equation}
where the efficiency factor $\epsilon \approx 0.05$.
 The total energy of gravitational waves is simply a sum of all the mentioned sources~\cite{Caprini:2015zlo}
\begin{equation}
\Omega_{\rm GW} h^2 (f) =\Omega h^2_{\rm col}(f) +\Omega h^2_{\rm sw} (f) +\Omega h^2_{\rm turb}(f).
\end{equation} 
%%%%%%%%%%%%%%%%%%%%%%%%%%%%%%%%%%%%%%%%%%%%%%%%%%%%%%%%%%%%%%%
%%%%%%%%%%%%%%%%%%%%%%%%%%%%%%%%%%%%%%%%%%%%%%%%%%%%%%%%%%%%%%%
%%%%%%%%%%%%%%%%%%%%%%%%%%%%%%%%%%%%%%%%%%%%%%%%%%%%%%%%%%%%%%%

\begin{figure}[t] 
	\centering
	\includegraphics[width=\textwidth]{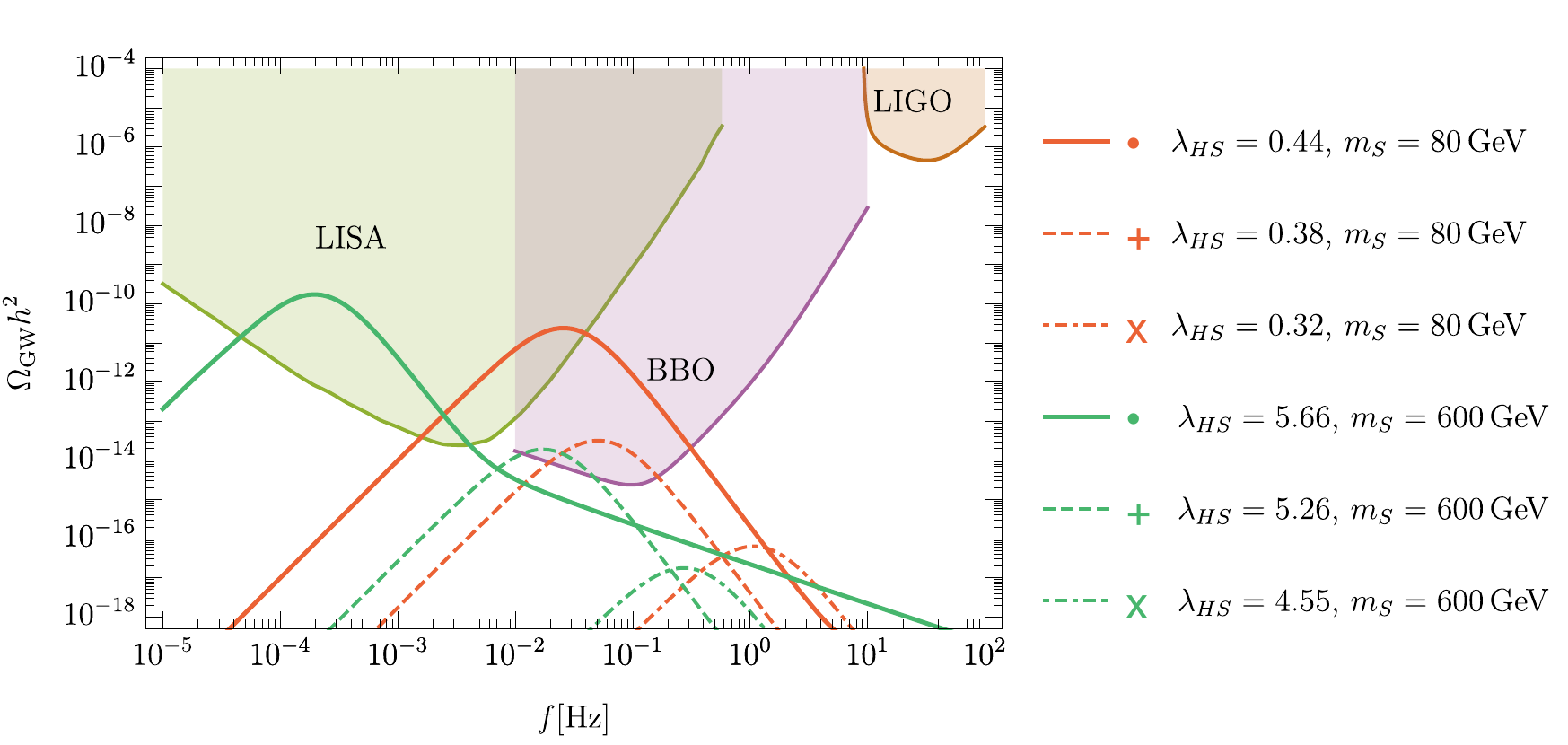} 
	\caption{Spectra of GWs from the electroweak phase transition for a few example points that are also marked in Fig.~\ref{fig:GWdetectionplot}. Projected sensitivities of the future based GW detectors such as LISA, BBO as well as current sensitivity of LIGO are also shown.}
	\label{fig:GWplot}
\end{figure}

Generally speaking, the magnitude of the GW signal grows with the strength of the phase transition. In Figs.~\ref{fig:GWdetectionplot} and \ref{fig:GWplot}, we illustrate this effect and show the regions of the parameter space accessible to the most promising future GW detectors, specifically LISA (with the most promising configuration A5M5) \cite{Bartolo:2016ami} and BBO \cite{Yagi:2011wg}. For comparison we also show the reach of LIGO \cite{TheLIGOScientific:2014jea} which cannot probe any part of the parameter space. 

For any value of $m_S$, the barrier separating the initial symmetric vacuum and the final EWSB one grows with $\lambda_{HS}$. This increases the VEV of the Higgs field after the transition, decreasing the transition temperature as tunnelling becomes suppressed, resulting in an initial unstable configuration of the field that survives longer due to the larger barrier. This means that the action $S_3$ of our solution grows while the transition temperature lowers. At some point, this inevitably leads to an over-suppression of the thermal tunnelling by a factor of $S_3/T$. This would cause the field to remain in the initial unstable configuration up until today ($T_* \approx 0$), which is of course excluded.

For very low transition temperatures, the vacuum decay is driven by quantum fluctuations and is only suppressed by the action $S_4$  \cite{Coleman:1977py} instead of $S_3/T$ in the exponent. The quantum tunnelling action $S_4$ still depends on the temperature since the potential does. However, this dependence is very weak as the potential is close to its zero temperature value when the quantum tunnelling becomes important. Calculation of the action is technically very similar to the procedure discussed in Section~\ref{sec:dynamicsofthephasetransition}. The important difference in this case is that our solution is four-dimensional, as it also includes the Euclidean time. Numerically, the resulting action is similar to the three-dimensional one and the decay probability is much smaller than in the thermally-induced decay case. In the end, this effect saves some part of the parameter space as the integrated decay probability is increased by adding this small probability to the integral between the temperature when quantum tunnelling dominates and $T_{\rm BBN}$. However, this is a subdominant effect and the part of the parameter space where it enables the phase transition to occur is negligible. Also, while the calculation of the GW signal in quite different in this case, the difference between vacuum energies in this case is still very large and the resulting signal magnitude would be just as large as in the high temperature case, allowing its observation up to the border of the allowed parameter space. 

An important point is that for all possible values of $m_S$, there is a significant region of model parameter space where a successful EWBG is followed by an observable GW signal. Specifically, for low masses, the coupling $\lambda_{HS}$ is too small for indirect detection at future based colliders (see e.g., Section~\ref{sec:collidersignals}) whereas the GW signal produced during the EWPT is within the reach of planned GW detectors. We can therefore conclude that the detection of GWs can be more a sensitive probe than the indirect collider searches. 

%%%%%%%%%%%%%%%%%%%%%%%%%%%%%%%%%%%%%%%%%%%%%%%%%%%%%%%%%%%%
%%%%%%%%%%%%%%%%%%%%%%%%%%%%%%%%%%%%%%%%%%%%%%%%%%%%%%%%%%%%%%%%%%%%%%
\subsection{Dark Matter signals}\label{sec:darkmatter}
In our simple model, the new scalar $S$ is stable and can serve as a DM candidate. One has to remember that all the DM considerations can become irrelevant if we extend the model with additional dark sector fields that couple to the SM only via the new scalar $S$ \cite{Silveira1985136,PhysRevD.50.3637,Burgess:2000yq,Alanne:2014bra}. This is possible because the new scalar would then be able to decay into light dark sector particles and thereby avoid all the DM detection limits. However, if the minimal model is realised, these bounds provide some of the strongest constraints on the allowed regions of the model parameter space (see e.g., Ref.~\cite{Beniwal:2015sdl}).

To calculate the abundance of $S$ in the universe today, we follow the standard analysis in Ref.~\cite{Gondolo:1990dk}. The Boltzmann equation has the form
\begin{equation}\label{eqn:abundeq}
\frac{d Y}{d x}=\frac{2\pi}{45}\frac{m_S^3}{x^4 H}\left(h_{\rm eff}+\frac{T}{3}\frac{d h_{\rm eff}}{d T} \right)\langle \sigma v \rangle \left( Y_{\rm eq}^2-Y^2 \right),
\end{equation} 
where $Y = n/s$, $x=m_S/T$, $\langle \sigma v \rangle$ is the thermally averaged annihilation cross section and $h_{\rm eff}$ is the effective number of entropy degrees of freedom. For the calculation of the $\sigma v$ into various SM final states, we use the results from Ref.~\cite{Cline:2013gha}. We numerically solve Eq.~\eqref{eqn:abundeq} and obtain the number density $n_0$ of the scalar $S$ today. Finally, the $S$ abundance/relic density is calculated using
\begin{equation}
\Omega_S h^2=\frac{m_S Y_0 s_0}{3 M_p^2 H^2_0} \sim  m_S Y_0 \times 2.76\times10^8. 
\end{equation}
Assuming that $S$ is the \emph{only} DM candidate, its relic density should match with the Planck measured value \cite{Ade:2015xua}
\begin{equation}\label{eqn:planck-2015}
	\Omega_{\rm DM}h^2 = 0.1188.	
\end{equation}
If one assumes a multicomponent dark sector, the $S$ abundance can be smaller but still cannot exceed that measured value not to overclose the universe.

To impose direct detection limits on the model parameter space, we calculate the spin-independent (SI) scalar-nucleon cross section \cite{Beniwal:2015sdl}
\begin{equation}\label{eqn:xsection}
\sigma_{\rm SI}=\frac{\lambda_{HS}^2 f_N^2}{4\pi}\frac{\mu^2 m_n^2}{m_S^2 m_h^4},
\end{equation} 
where $\mu=m_n m_S/(m_n+m_S)$ is the DM-nucleon reduced mass, $m_n=938.95$\mev and $f_N=0.3$ \cite{Cline:2013gha,Alarcon:2011zs,Alarcon:2012nr}. If the abundance of $S$ is smaller than the Planck measured value in Eq.~\eqref{eqn:planck-2015}, the SI cross section in Eq.~\eqref{eqn:xsection} must be appropriately scaled such that points with
\begin{equation}
\frac{\Omega_S}{\Omega_{\rm DM}}\sigma_{\rm SI} > \sigma_{\rm EXP}
\end{equation}
are excluded. 

\begin{figure}[t]
	\centering
	\includegraphics[width=\textwidth]{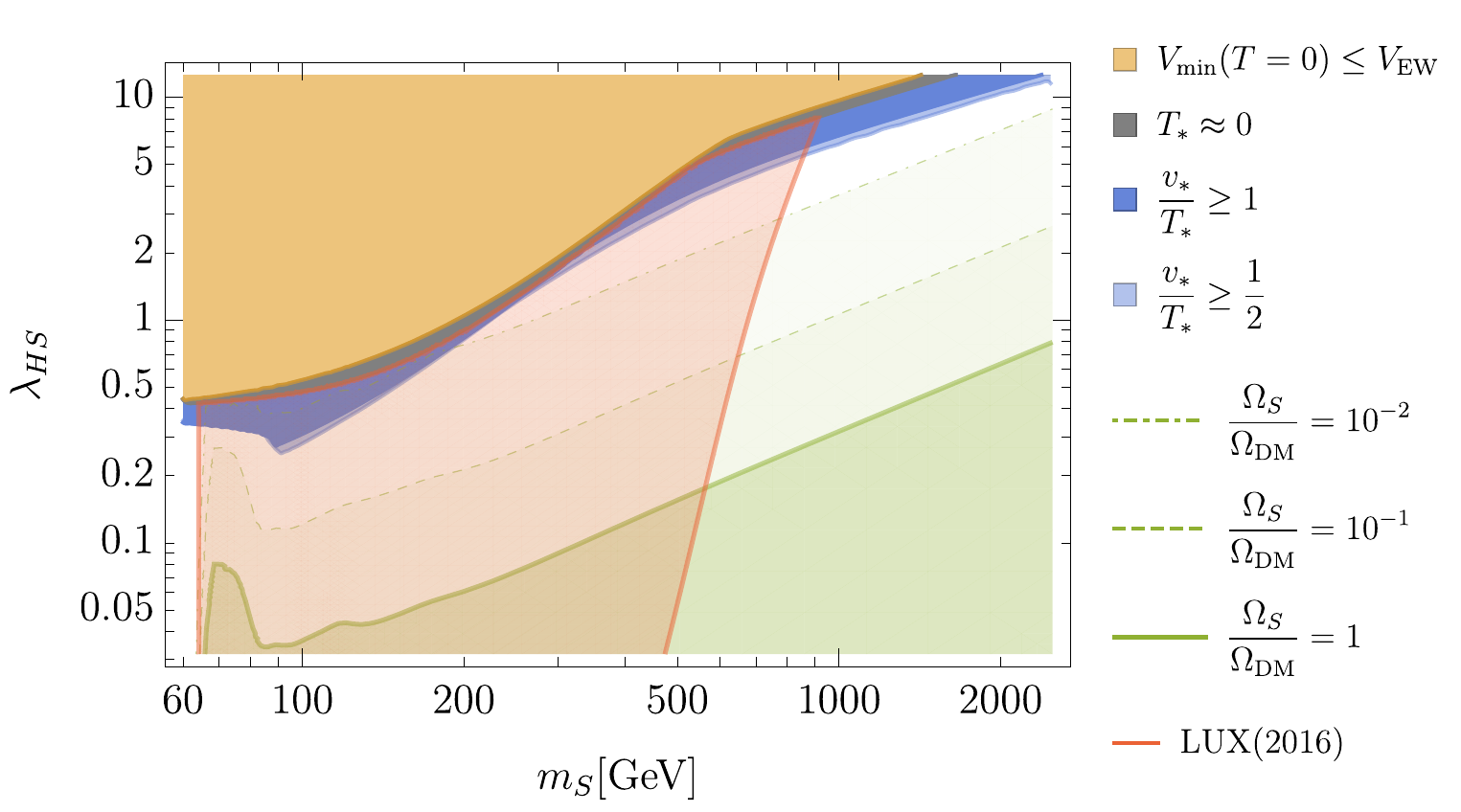} 
	\caption{Parameter space of the scalar singlet model relevant for EWBG together with the DM abundance and corresponding direct detection exclusion limits. Constraints from the vacuum structure of the theory are also taken into account, hence the reason why the abundance or the direct detection limits do not enter into the gray or yellow shaded regions.}
	\label{fig:DMplot}
\end{figure}

Currently, the strongest limits on the SI DM-nucleon cross section come from the LUX (2016) experiment \cite{Akerib:2016vxi}. Using these limits and the analysis presented above, we show the excluded regions of the model parameter space in Fig.~\ref{fig:DMplot}. In the regions where EWBG is possible, the scalar $S$ only accounts for less than $1\%$ of the total DM. However, the LUX (2016) experiment still severely constrains the model parameter space. Only a small portion of the model parameter space with a small $S$ abundance is still allowed, either requiring scalar masses $m_S>700$\gev or masses just above the Higgs resonance $m_S\sim m_h/2$.

Our results take into account the vacuum structure of the theory. The region where the EWPT does not occur is not constrained by these results because in that case, the freeze-out does not proceed in the electroweak vacuum; a much different calculation of the DM abundance would be required. However, to a large extent, this would be a pointless exercise since the region is excluded to begin with. There is only a small loop-hole to this argument, namely the region just below the no-EWPT excluded region, where the transition proceeds at very low temperature. This means that DM can freeze-out before EWPT occurs in a vacuum with $\langle S \rangle>0, \langle h \rangle=0$ for $m_S<600$\gev or $\langle S \rangle=\langle h \rangle=0$ for $m_S>600$\gev. Even this exotic possibility is mostly ruled out. For $m_S>600$\gev, DM freezes out in a vacuum at the origin with zero vacua for both scalars. This closes all the usual decay channels generated by the $hSS$ vertex since the Higgs VEV is now absent. The resulting DM abundance is higher than it would have been in the electroweak vacuum and this region is even more constrained by direct detection experiments. The $m_S<600$\gev region has to be considered in two parts. Firstly, in the $m_S<2m_h$ region where even though the $S$ VEV generates a $Shh$ vertex, the simple decay of $S$ into two Higgses is kinematically suppressed. As in the previous case, most of the usual decay channels via the Higgs decay into SM particles are closed due to a missing $hSS$ vertex. Only the $SS\rightarrow hh$ channel is available and as a result of the smaller cross section, a larger $S$ abundance is achieved which is again more constrained by direct detection searches. The last possibility is $2m_h<m_S<600$\gev, where the decay $S\rightarrow hh$ is possible. In principle, one could try to find points where the slightly larger DM abundance is depleted as $S$ decays into Higgses in a short time between its freeze-out and EWPT. Confirming this possibility would required a much more dedicated study in a negligibly small region of the parameter space. We will leave this possibility unresolved while saying only that even if such points exist, they are very fine-tuned and, indeed, their existence is open to question. 

%%%%%%%%%%%%%%%%%%%%%%%%%%%%%%%%%%%%%%%%%%%%%%%%%%%%%%%%%%%%%%%%%%%%%%
%%%%%%%%%%%%%%%%%%%%%%%%%%%%%%%%%%%%%%%%%%%%%%%%%%%%%%%%%%%%%%%%%%%%%

Our main conclusions are, firstly, $S$ cannot play the role of a single-component DM in regions where the EWBG is allowed. Secondly, in the regions where the EWBG is realised, even the very small remaining $S$ abundance is enough to generate severe constraints from direct detection null results. Moreover, the region of low mass and large coupling (and consequently a smaller relic density) that is usually considered as a hope for EWBG in this scenario is ruled out by the vacuum structure of the theory. However, it is important to keep in mind that adding a lighter particle (for e.g., a Dirac fermion \cite{Alanne:2014bra,Li:2014wia} or another scalar~\cite{Chao:2017vrq}) that couples only to the scalar $S$ can remove these constraints without affecting our predictions for EWBG. In that case, all the frozen-out scalars would simply decay into the dark sector particle which plays the role of a DM candidate. This is an important realisation as the scalar $S$ \emph{without} any new dark sector particles cannot account for all of the DM.  

%%%%%%%%%%%%%%%%%%%%%%%%%%%%%%%%%%%%%%%%%%%%%%%%%%%%%%%%%%%%%%%%%%%%%%
\section{Cosmological modification}\label{sec:cosmomod}
To ensure that we discuss all the parameter space where the scalar singlet model is viable, we also discuss a possible modification of the cosmological history which can expand this area significantly. We will focus on a very simple and generic cosmological modification that can describe the effects of most existing cosmological models.

We assume an additional contribution to the energy budget of the early universe $\rho_N$. The modified Friedmann equation reads
\begin{equation}\label{eq:friedmann}
H^2 \equiv \left( \frac{\dot{a}}{a} \right)^2 = \frac{8\pi}{3 M_{p}^2}\left( \frac{\rho_R}{a^4}+\frac{\rho_N}{a^n}\right),
\end{equation}
where $a \equiv a(t)$ is the scale factor and $n > 4$ such that the new component dilutes before it modifies any cosmological measurements. The first of such important measurements comes from Big Bang Nucleosynthesis (see e.g. Refs.~\cite{Cooke:2014,Agashe:2014kda}). We can directly measure the Hubble rate at that time since we precisely know when the neutrons have to freeze-out in order to save a fraction of them required to recreate observed abundances of light elements. While the observed expansion is consistent with a universe filled with the SM radiation, within experimental uncertainties, we can still add a small fraction of the additional component $\rho_N$.

First, we translate the effective number of neutrino species into a modification of the Hubble rate \cite{Simha:2008zj}
\begin{equation}
\left. \frac{H}{H_R} \right|_{\rm BBN}=\sqrt{1+\frac{7}{43}\Delta N_{\nu_{\rm eff}}},
\end{equation}
where $H_R$ is the standard case (i.e., SM radiation) and 
\begin{equation*}
	\Delta N_{\nu_{\rm eff}}= (N_{\nu_{\rm eff}} + 2\sigma)-N^{\rm SM}_{\nu_{\rm eff}}= (3.28+2\times0.28)-3.046 = 0.794	
\end{equation*}
is the difference between the effective number of neutrinos in the SM radiation case and the 2$\sigma$ experimental upper bound \cite{Cooke:2014,Agashe:2014kda}.

We also assume that the new component does not directly interact with the SM, so the usual relation between the scale factor and temperature holds
\begin{equation}\label{eq:rhorad}
\frac{\rho_{R}}{a^4}=\frac{\pi^2}{30}g T^4,
\end{equation} 
where $g$ is the number of degrees of freedom in the SM. This leads to a usual result for the Hubble rate in the radiation dominated case
\begin{equation}\label{eq:Hrad}
H_R=\sqrt{\frac{4\pi}{45}g}\frac{T^2}{M_p}.
\end{equation}

\begin{figure}[t]
	\centering
	\includegraphics[width=0.44\textwidth]{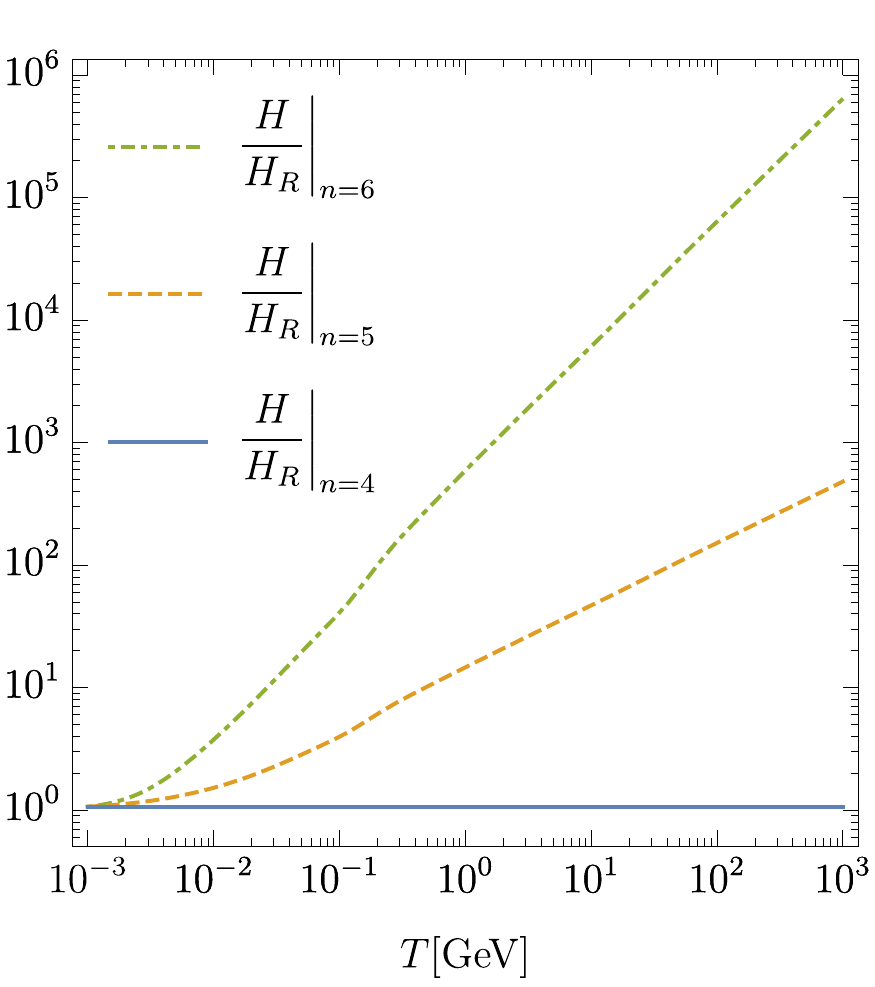} 
	\quad
	\includegraphics[width=0.5\textwidth]{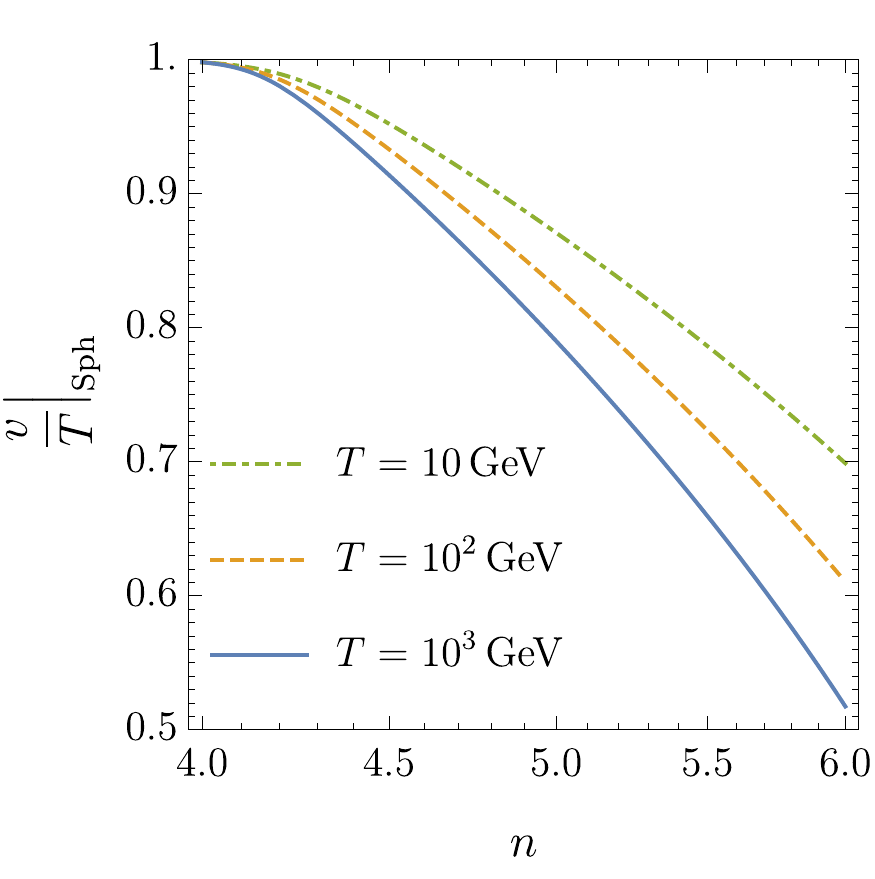}
	\caption{Left panel: Maximal modification of the Hubble rate $H$ that is not in conflict with any experimental bounds. Right panel: Values of the $v/T$ (at $T = T_*$) needed to avoid the washout of the baryon asymmetry after the EWPT as a function of the modification of the Hubble rate.}
	\label{fig:modcosmoplot}
\end{figure}

We are now ready to calculate an upper bound on the expansion rate at an earlier time (i.e., high temperature). The contribution from the new component grows quickly and dominates the total energy density. Once this occurs, we can neglect $\rho_R$ in Eq.~\eqref{eq:friedmann} and arrive at the following result
\begin{equation}\label{eq:Hmod}
\frac{H}{H_R} =
\sqrt{\left(\left. \frac{H}{H_R} \right|_{\rm BBN} \right)^2-1}
\left[\left(\frac{g}{g_{\rm BBN}}\right)^{\frac{1}{4}}\frac{T}{T_{\rm BBN}}\right]^{\frac{n-4}{2}},
\end{equation} 
where the values with the subscript ``BBN" are calculated at the BBN temperature $T_ {\rm BBN}=1$\mev and all the others are calculated at an earlier time corresponding to a temperature $T$. The resulting maximal modification is shown in the left panel of Fig.~\ref{fig:modcosmoplot}. The $n=6$ case can be realised in many cosmological models and results in an increase that can be as big as $10^5$ in temperatures around EWPT. This big modification has important consequences as discussed below. 

%%%%%%%%%%%%%%%%%%%%%%%%%%%%%%%%%%%%%%%%%%%%%%%%%%%%%%%%%%%%%%%%%%%%%%

%%%%%%%%%%%%%%%%%%%%%%%%%%%%%%%%%%%%%%%%%%%%%%%%%%%%%%%%%%%%%%%%%%%%%%
%%%%%%%%%%%%%%%%%%%%%%%%%%%%%%%%%%%%%%%%%%%%%%%%%%%%%%%%%%%%%%%%%%%%%
\subsection{The sphaleron bound and its cosmological modification}\label{sec:sphlaboundsmod}

We will now revisit the sphaleron bound discussed in Section~\ref{sec:baryo} and discuss its modification due to a non-standard cosmological history. The simplest criterion for decoupling of the sphalerons comes simply from requiring that the sphaleron rate is smaller than the Hubble rate after the EWPT
\begin{equation}
\Gamma_{\rm Sph} = T^4 \mathcal{B}_0 \frac{g}{4 \pi} \left(\frac{v}{T}\right)^7 \exp \left( -\frac{4 \pi}{g}\frac{v}{T} \right)\leq H,
\end{equation}
where the constant $\mathcal{B}_0$ encapsulate the details of the $SU(2)$ sphaleron calculation. Rigorous calculation of the value of $\mathcal{B}_0$ generally proves to be difficult and therefore a few different values have been used in the literature, leading to different bounds on $v/T$ \cite{Katz:2014bha,Quiros:1999jp,Funakubo:2009eg,Fuyuto:2014yia}. We use the value corresponding to the standard bound (see Eq.~\eqref{eqn:sphbound}) for the SM radiation dominated cosmology $H=H_R$. The right panel in Fig.~\ref{fig:modcosmoplot} shows the dependence of $v/T$ needed to decouple the sphalerons after the EWPT on the Hubble rate. Comparing both panels in Fig.~\ref{fig:modcosmoplot}, we see that the required ratio can be as low as $v/T \sim 1/2$ for cosmological models with $n=6$. We highlight this value in our results below to show the possible impact of such a cosmological modification.
%%%%%%%%%%%%%%%%%%%%%%%%%%%%%%%%%%%%%%%%%%%%%%%%%%%%%%%%%%%%%%%%%%%%%%
%%%%%%%%%%%%%%%%%%%%%%%%%%%%%%%%%%%%%%%%%%%%%%%%%%%%%%%%%%%%%%%%%%%%%%
\subsection{Modified cosmology and dark matter}

%%%%%%%%%%%%%%%%%%%%%%%%%%%%%%%%%%%%%%%%%%%%%%%%%%%%%%%%%%%%%%%%%%%%%%
\begin{figure}[t]
	\centering
	\includegraphics[width=\textwidth]{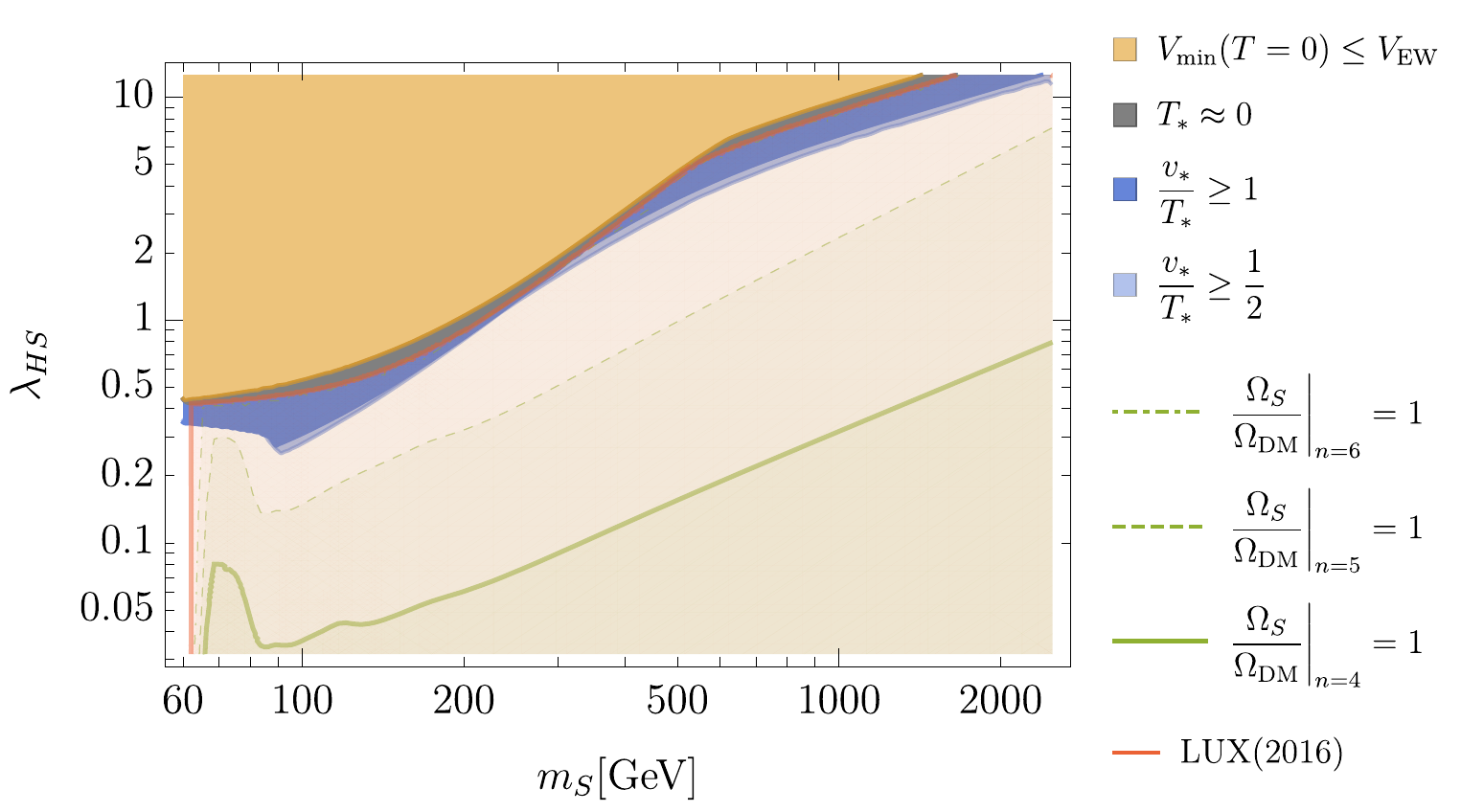}
	\caption{Parameter space of the scalar singlet model relevant for EWBG along with the DM abundance and direct detection constraints  due to cosmological modification. Values of $\lambda_{HS}$ along the green lines give the correct DM abundance for cosmological modification with given $n$. For $n=6$, the correct DM abundance can be achieved anywhere between the area excluded by the vacuum structure and the usual radiation domination case (i.e., $n=4$). The direct detection limits shown are based on the $S$ abundance with $n=6$.}
	\label{fig:DMmodplot}
\end{figure}

The abundance of $S$ is very sensitive to any modifications in the cosmological history. An increased Hubble rate $H$ will result in an earlier freeze-out of the scalar $S$ than in the standard case. This would leave a much larger abundance of $S$ in the universe today. To see this effect in action, we simply replace $H$ in Eq.~\eqref{eqn:abundeq} with a modified Hubble rate given in Eq.~\eqref{eq:Hmod}. The results are shown in Fig.~\ref{fig:DMmodplot}. As is evident, the $S$ abundance is increased by orders of magnitude. The $n=6$ case can achieve the correct abundance in all parts of the parameter space up to the region excluded by the vacuum structure. However, no parameter space that allows the correct DM abundance is opened up due to the modification. This is because the increased $S$ abundance also results in a more severe direct detection constraint.

It is still interesting to note that a higher expansion rate increases the DM abundance for large couplings while allowing smaller ones to be compatible for EWBG, thus bringing the two regions close together. 

%%%%%%%%%%%%%%%%%%%%%%%%%%%%%%%%%%%%%%%%%%%%%%%%%%%%%%%%%%%%%%%%%%%%
%%%%%%%%%%%%%%%%%%%%%%%%%%%%%%%%%%%%%%%%%%%%%%%%%%%%%%%%%%%%%%%%%%%%
\section{Conclusions}\label{sec:conclusions}

In this paper, we have studied the viability and detection prospects of a scalar singlet extension of the SM. We focused on two attractive features of this model, namely the possibility to facilitate EWBG and the DM candidate. We discussed various experimental probes of this scenario and their reach in parts of the model parameter space. These include collider signals, detection of GW from the phase transition and direct detection of DM. 

We studied the dynamics of the phase transition including the analysis of a region where a two-step phase transition occurs. In that case, the universe first transitions into a minimum along the $\langle h \rangle =0, \, \langle S \rangle>0$ vacuum configuration and then subsequently decays to the electroweak vacuum $\langle h \rangle = v_0, \langle S \rangle = 0$. This allowed us to accurately calculate the transition temperature and its strength, which in turn enabled us to predict the GW signals of the phase transition in all parts of the parameter space.

Our most important conclusion is that a significant portion of the model parameter space is accessible at the planned GW experiments but is beyond reach at the future collider experiments. 
The region of smaller coupling with the new scalar is especially attractive as it guarantees that our one-loop analysis is accurate and that no Landau poles are hit near the electroweak scale when the scalar coupling grows with the RG evolution (see Ref.~\cite{Curtin:2014jma}).

We also extensively tested the possibility that our new scalar is a DM candidate. Using the standard freeze-out of $S$ to compute its present relic density, we identified the model parameter space where $S$ satisfies the observed DM abundance and where it is constrained by the limits from the recent LUX (2016) experiment.

The conclusion here is that the correct DM abundance cannot be obtained simultaneously with a first-order EWPT. However, this small abundance is nevertheless enough to lead to exclusion by null results in direct dark matter search experiments. The situation is even worse in the region of small scalar mass and large coupling not constrained by direct detection experiments since it is all excluded by the vacuum structure. This is so because the universe in that case would not transition to the electroweak vacuum, at least not for perturbative values of the couplings $\lambda_S$ and $\lambda_{HS}$.

Only two small regions allowing EWBG remain viable, the first being the region close to the Higgs resonance $m_S\sim m_h/2$ and the another at scalar masses $m_S>700$\gev. However, it is important to remember that all the mentioned DM constraints can be circumvented if the scalar only serves as a mediator between a new DM candidate and the SM. The details of the EW phase transition in that case remain the same.

Lastly, we checked which part of the models parameter space opened up when the cosmological history is modified. To this end, we employed a simple cosmological model, assuming a new energy constituent that redshifts away faster than radiation (i.e., $\rho_N\propto a^{-n}$ for $n>4$). We placed bounds within which the expansion rate of the universe could be increased without spoiling any of the astrophysical observations.

We concluded that the cosmological modification has significant consequences for both EWBG and the DM abundance. This lead to a shift in the regions where the two can be matched with observations. However, an increase in the DM abundance will always be followed by severe constraints from direct detection experiments and as a result no new viable parameter space opens up. 

%%%%%%%%%%%%%%%%%%%%%%%%%%%%%%%%%%%%%%%%%%%%%%%%%%%%%%%%%%%%%%%%%%%%%%
%%%%%%%%%%%%%%%%%%%%%%%%%%%%%%%%%%%%%%%%%%%%%%%%%%%%%%%%%%%%%%%%%%%%%%
%%%%%%%%%%%%%%%%%%%%%%%%%%%%%%%%%%%%%%%%%%%%%%%%%%%%%%%%%%%%%%%%%%%%%%
%%%%%%%%%%%%%%%%%%%%%%%%%%%%%%%%%%%%%%%%%%%%%%%%%%%%%%%%%%%%%%
\section*{Acknowledgements}
This work was supported by the ARC Centre of Excellence for Particle Physics at the Terascale (CoEPP) (CE110001104) and the Centre for the Subatomic Structure of Matter (CSSM). JDW was supported in part by the U.S.\ Department of Energy (DE-SC0007859) and the Alexander von Humboldt Foundation (Humboldt Forschungspreis). ML was supported in part by the Polish National Science Centre under doctoral scholarship number 2015/16/T/ST2/00527 and MNiSW grant IP2015 043174. MW is supported by the Australian Research Council Future Fellowship FT140100244. 

%%%%%%%%%%%%%%%%%%%%%%%%%%%%%%%%%%%%%%%%%%%%%%%%%%%%%%%%%%%%%%%%%%%%%%
\appendix
\section{Effective potential}\label{sec:effpotappendix}
We use the following one-loop corrections to the zero temperature potential using the cutoff regularisation and on-shell scheme \cite{Curtin:2014jma,Delaunay:2007wb},
\begin{equation}\label{eqn:V1}
 V_{1-\rm loop}(h,S)=\sum_{i=h,\chi,W,Z,t,S}\frac{n_i}{64\pi^2}\left[m_{i}^4 \left( \log\frac{m^2_{i}}{m^2_{0 i}}-\frac{3}{2}\right)+2 m^2_{i} m^2_{0i}\right] ,
\end{equation}
where $n_{\{h,\chi,W,Z,t,S\}}=\{1,3,6,3,-12,1\}$ and $m_0$ are masses calculated at the electroweak VEV $S=0, \ h=v_0$. The field dependant masses are
\begin{equation}
\begin{split} 
m_W^2 =\frac{g^2}{4}h^2 &, \quad m_Z^2=\frac{g^2+g'^2}{4}h^2, \quad \quad m_t^2=\frac{y_t^2}{2}h^2, \\
m_\chi^2 & =-\mu^2+\lambda h^2+\lambda_{HS}S^2.
\end{split}
\end{equation}
The $h$ and $S$ masses are the eigenvalues of the following mixing matrix
\begin{equation}\label{eq:scalarmassmatrix}
M_{HS}=\begin{pmatrix}
-\mu^2+3\lambda h^2+\lambda_{HS}S^2 & 2\lambda_{HS} h S \\
2\lambda_{HS} h S & \mu_S^2+3\lambda_S S^2+\lambda_{HS}h^2
\end{pmatrix}.
\end{equation}
The finite temperature corrections are given by
\begin{equation}\label{eqn:VT}
 V_T(h,S,T)=\sum_{i= h,\chi,W,Z,S}\frac{n_iT^4}{2\pi^2} J_b\left(\frac{m^2_i}{T^2}\right)+\sum_{i= t}\frac{n_iT^4}{2\pi^2} J_f\left(\frac{m^2_i}{T^2}\right),
 \end{equation}
where 
\begin{equation}
J_{b/f }\left(\frac{m^2_i}{T^2}\right)=\int_0^\infty dk \, 
k^2\log\left[1\mp {\rm exp}\left(-\sqrt{\frac{k^2+m_i^2}{T^2}} \right) \right].
\end{equation}
The last important correction comes from resumming the multi-loop infrared divergent contributions to boson longitudinal polarisations \cite{Arnold:1992rz,Carrington:1991hz}. We achieve this by adding thermal corrections to scalars and longitudinal polarisations of the gauge bosons.
 
%\begin{equation}\label{eqn:Vr}
% V_r(h,S,T)=\sum_{i=h,\chi,W,Z,\gamma,S}\frac{\bar{n}_iT}{12\pi}\left[m^3_i(\phi)-\left(m^2_i+\Pi_i(T)\right)^{3/2}\right],
%\end{equation}
%where $\bar{n}_{\{h,\chi,W,Z,t\}}=\{1,3,2,1,1\}$.
The thermal corrections to masses can be obtained by expanding Eq.~\eqref{eqn:VT} to the leading order in $m^2/T^2$ \cite{Arnold:1992rz}. In our model, they are \cite{Curtin:2014jma}
\begin{equation}
\begin{split}
\Pi_h (T) & =\Pi_\chi(T)=T^2 \left( 
 \frac{g'^2}{16} + \frac{3 g}{16}+\frac{\lambda}{2}+\frac{y_t^2}{4} + \frac{\lambda_S}{12}
 \right),   \\
\Pi_S(T)& =T^2 \left(\frac{\lambda_{HS}}{3}+ \frac{\lambda_S}{4}  \right) , \quad
\Pi_W(T) =\frac{11}{6} g^2 T^2,  
%\Pi_Z(T)&=\frac{22}{3}\frac{(m^2_Z-m^2_W)}{v_0^2}T^2-m^2_W(\phi)\\
%\Pi_\gamma(T)&=m^2_W(\phi)+\frac{22}{3}\frac{m^2_W}{v_0^2}T^2.
\end{split}
\end{equation}
For the two scalars, the thermally corrected masses are the eigenvalues of the following mass matrix
\begin{equation}\label{eq:scalarmassmatrix}
M_{HS}+\begin{pmatrix}
\Pi_h (T)&0 \\
0  & \Pi_S (T)
\end{pmatrix},
\end{equation}
whereas the corrected masses of $Z$ and $\gamma$ (i.e., $m^2_{Z/\gamma}+\Pi_{Z/\gamma}(T)$) are the eigenvalues of the following mass matrix including thermal corrections
\begin{equation}
\begin{pmatrix}
\frac{1}{4} g^2 h^2+ \frac{11}{6} g^2T^2 & -\frac{1}{4}g' g h^2 \\[1.1mm]
-\frac{1}{4}g' g h^2 & \frac{1}{4} g'^2 h^2+\frac{11}{6} g'^2T^2
\end{pmatrix}.
\end{equation}
In all other cases, we can simply use the substitution
\begin{equation}
m_i^2 \rightarrow m_i^2 +\Pi_i.
\end{equation}
Our final estimate is simply a sum of the tree-level potential $V_{\mathrm{tree}}(h,S)$ in Eq.~\eqref{eqn:V0} and all of the above corrections
\begin{equation}
V_{\rm eff}(h,S,T)=V_{\textrm{tree}}(h,S)+V_{1-\rm loop}(h,S)+V_{T}(h,S,T).%+V_r(h,S,T).
\end{equation}

%%%%%%%%%%%%%%%%%%%%%%%%%%%%%%%%%%%%%%%%%%%%%%%%%%%%%%%%%%%%%%%%%%%%%%
%%%%%%%%%%%%%%%%%%%%%%%%%%%%%%%%%%%%%%%%%%%%%%%%%%%%%%%%%%%%%%%%%%%%%%
\bibliographystyle{JHEP}
\bibliography{NeutralScalarEWPTbibliography}

\providecommand{\href}[2]{#2}\begingroup\raggedright\begin{thebibliography}{10}

\bibitem{Aad:2012tfa}
{\scshape ATLAS} collaboration, G.~Aad et~al., \emph{{Observation of a new
  particle in the search for the Standard Model Higgs boson with the ATLAS
  detector at the LHC}},
  \href{http://dx.doi.org/10.1016/j.physletb.2012.08.020}{\emph{Phys. Lett.}
  {\bf B716} (2012) 1--29}, [\href{http://arxiv.org/abs/1207.7214}{{\tt
  1207.7214}}].

\bibitem{Chatrchyan:2012xdj}
{\scshape CMS} collaboration, S.~Chatrchyan et~al., \emph{{Observation of a new
  boson at a mass of 125 GeV with the CMS experiment at the LHC}},
  \href{http://dx.doi.org/10.1016/j.physletb.2012.08.021}{\emph{Phys. Lett.}
  {\bf B716} (2012) 30--61}, [\href{http://arxiv.org/abs/1207.7235}{{\tt
  1207.7235}}].

\bibitem{Abbott:2016blz}
{\scshape Virgo, LIGO Scientific} collaboration, B.~P. Abbott et~al.,
  \emph{{Observation of Gravitational Waves from a Binary Black Hole Merger}},
  \href{http://dx.doi.org/10.1103/PhysRevLett.116.061102}{\emph{Phys. Rev.
  Lett.} {\bf 116} (2016) 061102}, [\href{http://arxiv.org/abs/1602.03837}{{\tt
  1602.03837}}].

\bibitem{Kuzmin:1985mm}
V.~A. Kuzmin, V.~A. Rubakov and M.~E. Shaposhnikov, \emph{{On the Anomalous
  Electroweak Baryon Number Nonconservation in the Early Universe}},
  \href{http://dx.doi.org/10.1016/0370-2693(85)91028-7}{\emph{Phys. Lett.} {\bf
  B155} (1985) 36}.

\bibitem{Cohen:1993nk}
A.~G. Cohen, D.~B. Kaplan and A.~E. Nelson, \emph{{Progress in electroweak
  baryogenesis}},
  \href{http://dx.doi.org/10.1146/annurev.ns.43.120193.000331}{\emph{Ann. Rev.
  Nucl. Part. Sci.} {\bf 43} (1993) 27--70},
  [\href{http://arxiv.org/abs/hep-ph/9302210}{{\tt hep-ph/9302210}}].

\bibitem{Riotto:1999yt}
A.~Riotto and M.~Trodden, \emph{{Recent progress in baryogenesis}},
  \href{http://dx.doi.org/10.1146/annurev.nucl.49.1.35}{\emph{Ann. Rev. Nucl.
  Part. Sci.} {\bf 49} (1999) 35--75},
  [\href{http://arxiv.org/abs/hep-ph/9901362}{{\tt hep-ph/9901362}}].

\bibitem{Morrissey:2012db}
D.~E. Morrissey and M.~J. Ramsey-Musolf, \emph{{Electroweak baryogenesis}},
  \href{http://dx.doi.org/10.1088/1367-2630/14/12/125003}{\emph{New J. Phys.}
  {\bf 14} (2012) 125003}, [\href{http://arxiv.org/abs/1206.2942}{{\tt
  1206.2942}}].

\bibitem{Arnold:1992rz}
P.~B. Arnold and O.~Espinosa, \emph{{The Effective potential and first order
  phase transitions: Beyond leading-order}},
  \href{http://dx.doi.org/10.1103/PhysRevD.50.6662,
  10.1103/PhysRevD.47.3546}{\emph{Phys. Rev.} {\bf D47} (1993) 3546},
  [\href{http://arxiv.org/abs/hep-ph/9212235}{{\tt hep-ph/9212235}}].

\bibitem{Kajantie:1996qd}
K.~Kajantie, M.~Laine, K.~Rummukainen and M.~E. Shaposhnikov, \emph{{A
  Nonperturbative analysis of the finite T phase transition in SU(2) x U(1)
  electroweak theory}},
  \href{http://dx.doi.org/10.1016/S0550-3213(97)00164-8}{\emph{Nucl. Phys.}
  {\bf B493} (1997) 413--438},
  [\href{http://arxiv.org/abs/hep-lat/9612006}{{\tt hep-lat/9612006}}].

\bibitem{Curtin:2014jma}
D.~Curtin, P.~Meade and C.-T. Yu, \emph{{Testing Electroweak Baryogenesis with
  Future Colliders}},
  \href{http://dx.doi.org/10.1007/JHEP11(2014)127}{\emph{JHEP} {\bf 11} (2014)
  127}, [\href{http://arxiv.org/abs/1409.0005}{{\tt 1409.0005}}].

\bibitem{Kotwal:2016tex}
A.~V. Kotwal, M.~J. Ramsey-Musolf, J.~M. No and P.~Winslow,
  \emph{{Singlet-catalyzed electroweak phase transitions in the 100 TeV
  frontier}}, \href{http://dx.doi.org/10.1103/PhysRevD.94.035022}{\emph{Phys.
  Rev.} {\bf D94} (2016) 035022}, [\href{http://arxiv.org/abs/1605.06123}{{\tt
  1605.06123}}].

\bibitem{Damgaard:2015con}
P.~H. Damgaard, A.~Haarr, D.~O'Connell and A.~Tranberg, \emph{{Effective Field
  Theory and Electroweak Baryogenesis in the Singlet-Extended Standard Model}},
  \href{http://dx.doi.org/10.1007/JHEP02(2016)107}{\emph{JHEP} {\bf 02} (2016)
  107}, [\href{http://arxiv.org/abs/1512.01963}{{\tt 1512.01963}}].

\bibitem{Damgaard:2013kva}
P.~H. Damgaard, D.~O'Connell, T.~C. Petersen and A.~Tranberg,
  \emph{{Constraints on New Physics from Baryogenesis and Large Hadron Collider
  Data}}, \href{http://dx.doi.org/10.1103/PhysRevLett.111.221804}{\emph{Phys.
  Rev. Lett.} {\bf 111} (2013) 221804},
  [\href{http://arxiv.org/abs/1305.4362}{{\tt 1305.4362}}].

\bibitem{Brauner:2016fla}
T.~Brauner, T.~V.~I. Tenkanen, A.~Tranberg, A.~Vuorinen and D.~J. Weir,
  \emph{{Dimensional reduction of the Standard Model coupled to a new singlet
  scalar field}},  \href{http://arxiv.org/abs/1609.06230}{{\tt 1609.06230}}.

\bibitem{Silveira1985136}
V.~Silveira and A.~Zee, \emph{Scalar phantoms},
  \href{http://dx.doi.org/http://dx.doi.org/10.1016/0370-2693(85)90624-0}{\emph{Phys.
  Lett.} {\bf 161B} (1985) 136}.

\bibitem{PhysRevD.50.3637}
J.~McDonald, \emph{Gauge singlet scalars as cold dark matter},
  \href{http://dx.doi.org/10.1103/PhysRevD.50.3637}{\emph{Phys. Rev. D} {\bf
  50} (1994) 3637}.

\bibitem{Burgess:2000yq}
C.~P. Burgess, M.~Pospelov and T.~ter Veldhuis, \emph{{The Minimal model of
  nonbaryonic dark matter: A singlet scalar}},
  \href{http://dx.doi.org/10.1016/S0550-3213(01)00513-2}{\emph{Nucl. Phys.}
  {\bf B619} (2001) 709--728}, [\href{http://arxiv.org/abs/hep-ph/0011335}{{\tt
  hep-ph/0011335}}].

\bibitem{Katz:2014bha}
A.~Katz and M.~Perelstein, \emph{{Higgs Couplings and Electroweak Phase
  Transition}}, \href{http://dx.doi.org/10.1007/JHEP07(2014)108}{\emph{JHEP}
  {\bf 07} (2014) 108}, [\href{http://arxiv.org/abs/1401.1827}{{\tt
  1401.1827}}].

\bibitem{Grojean:2006bp}
C.~Grojean and G.~Servant, \emph{{Gravitational Waves from Phase Transitions at
  the Electroweak Scale and Beyond}},
  \href{http://dx.doi.org/10.1103/PhysRevD.75.043507}{\emph{Phys. Rev.} {\bf
  D75} (2007) 043507}, [\href{http://arxiv.org/abs/hep-ph/0607107}{{\tt
  hep-ph/0607107}}].

\bibitem{Artymowski:2016tme}
M.~Artymowski, M.~Lewicki and J.~D. Wells, \emph{{Gravitational wave and
  collider implications of electroweak baryogenesis aided by non-standard
  cosmology}},  \href{http://arxiv.org/abs/1609.07143}{{\tt 1609.07143}}.

\bibitem{Vaskonen:2016yiu}
V.~Vaskonen, \emph{{Electroweak baryogenesis and gravitational waves from a
  real scalar singlet}},  \href{http://arxiv.org/abs/1611.02073}{{\tt
  1611.02073}}.

\bibitem{Ashoorioon:2009nf}
A.~Ashoorioon and T.~Konstandin, \emph{{Strong electroweak phase transitions
  without collider traces}},
  \href{http://dx.doi.org/10.1088/1126-6708/2009/07/086}{\emph{JHEP} {\bf 07}
  (2009) 086}, [\href{http://arxiv.org/abs/0904.0353}{{\tt 0904.0353}}].

\bibitem{Enqvist:2014zqa}
K.~Enqvist, S.~Nurmi, T.~Tenkanen and K.~Tuominen, \emph{{Standard Model with a
  real singlet scalar and inflation}},
  \href{http://dx.doi.org/10.1088/1475-7516/2014/08/035}{\emph{JCAP} {\bf 1408}
  (2014) 035}, [\href{http://arxiv.org/abs/1407.0659}{{\tt 1407.0659}}].

\bibitem{Tenkanen:2016idg}
T.~Tenkanen, K.~Tuominen and V.~Vaskonen, \emph{{A Strong Electroweak Phase
  Transition from the Inflaton Field}},
  \href{http://dx.doi.org/10.1088/1475-7516/2016/09/037}{\emph{JCAP} {\bf 1609}
  (2016) 037}, [\href{http://arxiv.org/abs/1606.06063}{{\tt 1606.06063}}].

\bibitem{Huang:2016cjm}
P.~Huang, A.~J. Long and L.-T. Wang, \emph{{Probing the Electroweak Phase
  Transition with Higgs Factories and Gravitational Waves}},
  \href{http://dx.doi.org/10.1103/PhysRevD.94.075008}{\emph{Phys. Rev.} {\bf
  D94} (2016) 075008}, [\href{http://arxiv.org/abs/1608.06619}{{\tt
  1608.06619}}].

\bibitem{Kakizaki:2015wua}
M.~Kakizaki, S.~Kanemura and T.~Matsui, \emph{{Gravitational waves as a probe
  of extended scalar sectors with the first order electroweak phase
  transition}}, \href{http://dx.doi.org/10.1103/PhysRevD.92.115007}{\emph{Phys.
  Rev.} {\bf D92} (2015) 115007}, [\href{http://arxiv.org/abs/1509.08394}{{\tt
  1509.08394}}].

\bibitem{Hashino:2016rvx}
K.~Hashino, M.~Kakizaki, S.~Kanemura and T.~Matsui, \emph{{Synergy between
  measurements of gravitational waves and the triple-Higgs coupling in probing
  the first-order electroweak phase transition}},
  \href{http://dx.doi.org/10.1103/PhysRevD.94.015005}{\emph{Phys. Rev.} {\bf
  D94} (2016) 015005}, [\href{http://arxiv.org/abs/1604.02069}{{\tt
  1604.02069}}].

\bibitem{Hashino:2016xoj}
K.~Hashino, M.~Kakizaki, S.~Kanemura, P.~Ko and T.~Matsui, \emph{{Gravitational
  waves and Higgs boson couplings for exploring first order phase transition in
  the model with a singlet scalar field}},
  \href{http://dx.doi.org/10.1016/j.physletb.2016.12.052}{\emph{Phys. Lett.}
  {\bf B766} (2017) 49--54}, [\href{http://arxiv.org/abs/1609.00297}{{\tt
  1609.00297}}].

\bibitem{Kobakhidze:2016mch}
A.~Kobakhidze, A.~Manning and J.~Yue, \emph{{Gravitational Waves from the Phase
  Transition of a Non-linearly Realised Electroweak Gauge Symmetry}},
  \href{http://arxiv.org/abs/1607.00883}{{\tt 1607.00883}}.

\bibitem{Chala:2016ykx}
M.~Chala, G.~Nardini and I.~Sobolev, \emph{{Unified explanation for dark matter
  and electroweak baryogenesis with direct detection and gravitational wave
  signatures}}, \href{http://dx.doi.org/10.1103/PhysRevD.94.055006}{\emph{Phys.
  Rev.} {\bf D94} (2016) 055006}, [\href{http://arxiv.org/abs/1605.08663}{{\tt
  1605.08663}}].

\bibitem{Choi:1993cv}
J.~Choi and R.~R. Volkas, \emph{{Real Higgs singlet and the electroweak phase
  transition in the Standard Model}},
  \href{http://dx.doi.org/10.1016/0370-2693(93)91013-D}{\emph{Phys. Lett.} {\bf
  B317} (1993) 385--391}, [\href{http://arxiv.org/abs/hep-ph/9308234}{{\tt
  hep-ph/9308234}}].

\bibitem{Dev:2016feu}
P.~S.~B. Dev and A.~Mazumdar, \emph{{Probing the Scale of New Physics by
  Advanced LIGO/VIRGO}},
  \href{http://dx.doi.org/10.1103/PhysRevD.93.104001}{\emph{Phys. Rev.} {\bf
  D93} (2016) 104001}, [\href{http://arxiv.org/abs/1602.04203}{{\tt
  1602.04203}}].

\bibitem{Balazs:2016tbi}
C.~Balazs, A.~Fowlie, A.~Mazumdar and G.~White, \emph{{Gravitational waves at
  aLIGO and vacuum stability with a scalar singlet extension of the Standard
  Model}}, \href{http://dx.doi.org/10.1103/PhysRevD.95.043505}{\emph{Phys.
  Rev.} {\bf D95} (2017) 043505}, [\href{http://arxiv.org/abs/1611.01617}{{\tt
  1611.01617}}].

\bibitem{Khan:2014kba}
N.~Khan and S.~Rakshit, \emph{{Study of electroweak vacuum metastability with a
  singlet scalar dark matter}},
  \href{http://dx.doi.org/10.1103/PhysRevD.90.113008}{\emph{Phys. Rev.} {\bf
  D90} (2014) 113008}, [\href{http://arxiv.org/abs/1407.6015}{{\tt
  1407.6015}}].

\bibitem{Cline:2013gha}
J.~M. Cline, K.~Kainulainen, P.~Scott and C.~Weniger, \emph{{Update on scalar
  singlet dark matter}}, \href{http://dx.doi.org/10.1103/PhysRevD.92.039906,
  10.1103/PhysRevD.88.055025}{\emph{Phys. Rev.} {\bf D88} (2013) 055025},
  [\href{http://arxiv.org/abs/1306.4710}{{\tt 1306.4710}}].

\bibitem{Beniwal:2015sdl}
A.~Beniwal, F.~Rajec, C.~Savage, P.~Scott, C.~Weniger, M.~White et~al.,
  \emph{{Combined analysis of effective Higgs portal dark matter models}},
  \href{http://dx.doi.org/10.1103/PhysRevD.93.115016}{\emph{Phys. Rev.} {\bf
  D93} (2016) 115016}, [\href{http://arxiv.org/abs/1512.06458}{{\tt
  1512.06458}}].

\bibitem{He:2016mls}
X.-G. He and J.~Tandean, \emph{{New LUX and PandaX-II Results Illuminating the
  Simplest Higgs-Portal Dark Matter Models}},
  \href{http://dx.doi.org/10.1007/JHEP12(2016)074}{\emph{JHEP} {\bf 12} (2016)
  074}, [\href{http://arxiv.org/abs/1609.03551}{{\tt 1609.03551}}].

\bibitem{Joyce:1996cp}
M.~Joyce, \emph{{Electroweak Baryogenesis and the Expansion Rate of the
  Universe}}, \href{http://dx.doi.org/10.1103/PhysRevD.55.1875}{\emph{Phys.
  Rev.} {\bf D55} (1997) 1875--1878},
  [\href{http://arxiv.org/abs/hep-ph/9606223}{{\tt hep-ph/9606223}}].

\bibitem{Joyce:1997fc}
M.~Joyce and T.~Prokopec, \emph{{Turning around the sphaleron bound:
  Electroweak baryogenesis in an alternative postinflationary cosmology}},
  \href{http://dx.doi.org/10.1103/PhysRevD.57.6022}{\emph{Phys. Rev.} {\bf D57}
  (1998) 6022--6049}, [\href{http://arxiv.org/abs/hep-ph/9709320}{{\tt
  hep-ph/9709320}}].

\bibitem{Servant:2001jh}
G.~Servant, \emph{{A Way to reopen the window for electroweak baryogenesis}},
  \href{http://dx.doi.org/10.1088/1126-6708/2002/01/044}{\emph{JHEP} {\bf 01}
  (2002) 044}, [\href{http://arxiv.org/abs/hep-ph/0112209}{{\tt
  hep-ph/0112209}}].

\bibitem{Lewicki:2016efe}
M.~Lewicki, T.~Rindler-Daller and J.~D. Wells, \emph{{Enabling Electroweak
  Baryogenesis through Dark Matter}},
  \href{http://dx.doi.org/10.1007/JHEP06(2016)055}{\emph{JHEP} {\bf 06} (2016)
  055}, [\href{http://arxiv.org/abs/1601.01681}{{\tt 1601.01681}}].

\bibitem{Lewicki:2016oqx}
M.~Lewicki, \emph{{EW baryogenesis via DM}},  in \emph{{28th Rencontres de
  Blois on Particle Physics and Cosmology Blois, France, May 29-June 3, 2016}},
  2016.
\newblock \href{http://arxiv.org/abs/1611.02387}{{\tt 1611.02387}}.

\bibitem{Quiros:1999jp}
M.~Quiros, \emph{{Finite temperature field theory and phase transitions}},  in
  \emph{{High energy physics and cosmology. Proceedings, Summer School,
  Trieste, Italy, June 29-July 17, 1998}}, pp.~187--259, 1999.
\newblock \href{http://arxiv.org/abs/hep-ph/9901312}{{\tt hep-ph/9901312}}.

\bibitem{Funakubo:2009eg}
K.~Funakubo and E.~Senaha, \emph{{Electroweak phase transition, critical
  bubbles and sphaleron decoupling condition in the MSSM}},
  \href{http://dx.doi.org/10.1103/PhysRevD.79.115024}{\emph{Phys. Rev.} {\bf
  D79} (2009) 115024}, [\href{http://arxiv.org/abs/0905.2022}{{\tt
  0905.2022}}].

\bibitem{Fuyuto:2014yia}
K.~Fuyuto and E.~Senaha, \emph{{Improved sphaleron decoupling condition and the
  Higgs coupling constants in the real singlet-extended standard model}},
  \href{http://dx.doi.org/10.1103/PhysRevD.90.015015}{\emph{Phys. Rev.} {\bf
  D90} (2014) 015015}, [\href{http://arxiv.org/abs/1406.0433}{{\tt
  1406.0433}}].

\bibitem{Linde:1981zj}
A.~D. Linde, \emph{{Decay of the False Vacuum at Finite Temperature}},
  \href{http://dx.doi.org/10.1016/0550-3213(83)90293-6}{\emph{Nucl. Phys.} {\bf
  B216} (1983) 421}.

\bibitem{Linde:1980tt}
A.~D. Linde, \emph{{Fate of the False Vacuum at Finite Temperature: Theory and
  Applications}},
  \href{http://dx.doi.org/10.1016/0370-2693(81)90281-1}{\emph{Phys. Lett.} {\bf
  B100} (1981) 37}.

\bibitem{Cline:1999wi}
J.~M. Cline, G.~D. Moore and G.~Servant, \emph{{Was the electroweak phase
  transition preceded by a color broken phase?}},
  \href{http://dx.doi.org/10.1103/PhysRevD.60.105035}{\emph{Phys. Rev.} {\bf
  D60} (1999) 105035}, [\href{http://arxiv.org/abs/hep-ph/9902220}{{\tt
  hep-ph/9902220}}].

\bibitem{Profumo:2010kp}
S.~Profumo, L.~Ubaldi and C.~Wainwright, \emph{{Singlet Scalar Dark Matter:
  monochromatic gamma rays and metastable vacua}},
  \href{http://dx.doi.org/10.1103/PhysRevD.82.123514}{\emph{Phys. Rev.} {\bf
  D82} (2010) 123514}, [\href{http://arxiv.org/abs/1009.5377}{{\tt
  1009.5377}}].

\bibitem{Wainwright:2011kj}
C.~L. Wainwright, \emph{{CosmoTransitions: Computing Cosmological Phase
  Transition Temperatures and Bubble Profiles with Multiple Fields}},
  \href{http://dx.doi.org/10.1016/j.cpc.2012.04.004}{\emph{Comput. Phys.
  Commun.} {\bf 183} (2012) 2006--2013},
  [\href{http://arxiv.org/abs/1109.4189}{{\tt 1109.4189}}].

\bibitem{Belanger:2014vza}
G.~Bélanger, F.~Boudjema, A.~Pukhov and A.~Semenov, \emph{{micrOMEGAs4.1: two
  dark matter candidates}},
  \href{http://dx.doi.org/10.1016/j.cpc.2015.03.003}{\emph{Comput. Phys.
  Commun.} {\bf 192} (2015) 322--329},
  [\href{http://arxiv.org/abs/1407.6129}{{\tt 1407.6129}}].

\bibitem{Goertz:2013kp}
F.~Goertz, A.~Papaefstathiou, L.~L. Yang and J.~Zurita, \emph{{Higgs Boson
  self-coupling measurements using ratios of cross sections}},
  \href{http://dx.doi.org/10.1007/JHEP06(2013)016}{\emph{JHEP} {\bf 06} (2013)
  016}, [\href{http://arxiv.org/abs/1301.3492}{{\tt 1301.3492}}].

\bibitem{Asner:2013psa}
D.~M. Asner et~al., \emph{{ILC Higgs White Paper}},  in \emph{{Proceedings,
  Community Summer Study 2013: Snowmass on the Mississippi (CSS2013):
  Minneapolis, MN, USA, July 29-August 6, 2013}}, 2013.
\newblock \href{http://arxiv.org/abs/1310.0763}{{\tt 1310.0763}}.

\bibitem{Contino:2016spe}
R.~Contino et~al., \emph{{Physics at a 100 TeV pp collider: Higgs and EW
  symmetry breaking studies}},  \href{http://arxiv.org/abs/1606.09408}{{\tt
  1606.09408}}.

\bibitem{Barr:2014sga}
A.~J. Barr, M.~J. Dolan, C.~Englert, D.~E. Ferreira~de Lima and M.~Spannowsky,
  \emph{{Higgs Self-Coupling Measurements at a 100 TeV Hadron Collider}},
  \href{http://dx.doi.org/10.1007/JHEP02(2015)016}{\emph{JHEP} {\bf 02} (2015)
  016}, [\href{http://arxiv.org/abs/1412.7154}{{\tt 1412.7154}}].

\bibitem{Englert:2013tya}
C.~Englert and M.~McCullough, \emph{{Modified Higgs Sectors and NLO Associated
  Production}}, \href{http://dx.doi.org/10.1007/JHEP07(2013)168}{\emph{JHEP}
  {\bf 07} (2013) 168}, [\href{http://arxiv.org/abs/1303.1526}{{\tt
  1303.1526}}].

\bibitem{Dawson:2013bba}
S.~Dawson et~al., \emph{{Working Group Report: Higgs Boson}},  in
  \emph{{Proceedings, 2013 Community Summer Study on the Future of U.S.
  Particle Physics: Snowmass on the Mississippi (CSS2013): Minneapolis, MN,
  USA, July 29-August 6, 2013}}, 2013.
\newblock \href{http://arxiv.org/abs/1310.8361}{{\tt 1310.8361}}.

\bibitem{Kamionkowski:1993fg}
M.~Kamionkowski, A.~Kosowsky and M.~S. Turner, \emph{{Gravitational radiation
  from first order phase transitions}},
  \href{http://dx.doi.org/10.1103/PhysRevD.49.2837}{\emph{Phys. Rev.} {\bf D49}
  (1994) 2837--2851}, [\href{http://arxiv.org/abs/astro-ph/9310044}{{\tt
  astro-ph/9310044}}].

\bibitem{Huber:2008hg}
S.~J. Huber and T.~Konstandin, \emph{{Gravitational Wave Production by
  Collisions: More Bubbles}},
  \href{http://dx.doi.org/10.1088/1475-7516/2008/09/022}{\emph{JCAP} {\bf 0809}
  (2008) 022}, [\href{http://arxiv.org/abs/0806.1828}{{\tt 0806.1828}}].

\bibitem{Jinno:2016vai}
R.~Jinno and M.~Takimoto, \emph{{Gravitational waves from bubble collisions:
  analytic derivation}},
  \href{http://dx.doi.org/10.1103/PhysRevD.95.024009}{\emph{Phys. Rev.} {\bf
  D95} (2017) 024009}, [\href{http://arxiv.org/abs/1605.01403}{{\tt
  1605.01403}}].

\bibitem{Hindmarsh:2013xza}
M.~Hindmarsh, S.~J. Huber, K.~Rummukainen and D.~J. Weir, \emph{{Gravitational
  waves from the sound of a first order phase transition}},
  \href{http://dx.doi.org/10.1103/PhysRevLett.112.041301}{\emph{Phys. Rev.
  Lett.} {\bf 112} (2014) 041301}, [\href{http://arxiv.org/abs/1304.2433}{{\tt
  1304.2433}}].

\bibitem{Hindmarsh:2015qta}
M.~Hindmarsh, S.~J. Huber, K.~Rummukainen and D.~J. Weir, \emph{{Numerical
  simulations of acoustically generated gravitational waves at a first order
  phase transition}},
  \href{http://dx.doi.org/10.1103/PhysRevD.92.123009}{\emph{Phys. Rev.} {\bf
  D92} (2015) 123009}, [\href{http://arxiv.org/abs/1504.03291}{{\tt
  1504.03291}}].

\bibitem{Caprini:2009yp}
C.~Caprini, R.~Durrer and G.~Servant, \emph{{The stochastic gravitational wave
  background from turbulence and magnetic fields generated by a first-order
  phase transition}},
  \href{http://dx.doi.org/10.1088/1475-7516/2009/12/024}{\emph{JCAP} {\bf 0912}
  (2009) 024}, [\href{http://arxiv.org/abs/0909.0622}{{\tt 0909.0622}}].

\bibitem{Caprini:2015zlo}
C.~Caprini et~al., \emph{{Science with the space-based interferometer eLISA.
  II: Gravitational waves from cosmological phase transitions}},
  \href{http://dx.doi.org/10.1088/1475-7516/2016/04/001}{\emph{JCAP} {\bf 1604}
  (2016) 001}, [\href{http://arxiv.org/abs/1512.06239}{{\tt 1512.06239}}].

\bibitem{Espinosa:2010hh}
J.~R. Espinosa, T.~Konstandin, J.~M. No and G.~Servant, \emph{{Energy Budget of
  Cosmological First-order Phase Transitions}},
  \href{http://dx.doi.org/10.1088/1475-7516/2010/06/028}{\emph{JCAP} {\bf 1006}
  (2010) 028}, [\href{http://arxiv.org/abs/1004.4187}{{\tt 1004.4187}}].

\bibitem{Bodeker:2009qy}
D.~Bodeker and G.~D. Moore, \emph{{Can electroweak bubble walls run away?}},
  \href{http://dx.doi.org/10.1088/1475-7516/2009/05/009}{\emph{JCAP} {\bf 0905}
  (2009) 009}, [\href{http://arxiv.org/abs/0903.4099}{{\tt 0903.4099}}].

\bibitem{Megevand:2009gh}
A.~Megevand and A.~D. Sanchez, \emph{{Velocity of electroweak bubble walls}},
  \href{http://dx.doi.org/10.1016/j.nuclphysb.2009.09.019}{\emph{Nucl. Phys.}
  {\bf B825} (2010) 151--176}, [\href{http://arxiv.org/abs/0908.3663}{{\tt
  0908.3663}}].

\bibitem{Bartolo:2016ami}
N.~Bartolo et~al., \emph{{Science with the space-based interferometer LISA. IV:
  Probing inflation with gravitational waves}},
  \href{http://dx.doi.org/10.1088/1475-7516/2016/12/026}{\emph{JCAP} {\bf 1612}
  (2016) 026}, [\href{http://arxiv.org/abs/1610.06481}{{\tt 1610.06481}}].

\bibitem{Yagi:2011wg}
K.~Yagi and N.~Seto, \emph{{Detector configuration of DECIGO/BBO and
  identification of cosmological neutron-star binaries}},
  \href{http://dx.doi.org/10.1103/PhysRevD.83.044011}{\emph{Phys. Rev.} {\bf
  D83} (2011) 044011}, [\href{http://arxiv.org/abs/1101.3940}{{\tt
  1101.3940}}].

\bibitem{TheLIGOScientific:2014jea}
{\scshape LIGO Scientific} collaboration, J.~Aasi et~al., \emph{{Advanced
  LIGO}}, \href{http://dx.doi.org/10.1088/0264-9381/32/7/074001}{\emph{Class.
  Quant. Grav.} {\bf 32} (2015) 074001},
  [\href{http://arxiv.org/abs/1411.4547}{{\tt 1411.4547}}].

\bibitem{Coleman:1977py}
S.~R. Coleman, \emph{{The Fate of the False Vacuum. 1. Semiclassical Theory}},
  \href{http://dx.doi.org/10.1103/PhysRevD.15.2929,
  10.1103/PhysRevD.16.1248}{\emph{Phys. Rev.} {\bf D15} (1977) 2929--2936}.

\bibitem{Alanne:2014bra}
T.~Alanne, K.~Tuominen and V.~Vaskonen, \emph{{Strong phase transition, dark
  matter and vacuum stability from simple hidden sectors}},
  \href{http://dx.doi.org/10.1016/j.nuclphysb.2014.11.001}{\emph{Nucl. Phys.}
  {\bf B889} (2014) 692--711}, [\href{http://arxiv.org/abs/1407.0688}{{\tt
  1407.0688}}].

\bibitem{Gondolo:1990dk}
P.~Gondolo and G.~Gelmini, \emph{{Cosmic abundances of stable particles:
  Improved analysis}},
  \href{http://dx.doi.org/10.1016/0550-3213(91)90438-4}{\emph{Nucl. Phys.} {\bf
  B360} (1991) 145--179}.

\bibitem{Ade:2015xua}
{\scshape Planck} collaboration, P.~A.~R. Ade et~al., \emph{{Planck 2015
  results. XIII. Cosmological parameters}},
  \href{http://dx.doi.org/10.1051/0004-6361/201525830}{\emph{Astron.
  Astrophys.} {\bf 594} (2016) A13},
  [\href{http://arxiv.org/abs/1502.01589}{{\tt 1502.01589}}].

\bibitem{Alarcon:2011zs}
J.~M. Alarcon, J.~Martin~Camalich and J.~A. Oller, \emph{{The chiral
  representation of the $\pi N$ scattering amplitude and the pion-nucleon sigma
  term}}, \href{http://dx.doi.org/10.1103/PhysRevD.85.051503}{\emph{Phys. Rev.}
  {\bf D85} (2012) 051503}, [\href{http://arxiv.org/abs/1110.3797}{{\tt
  1110.3797}}].

\bibitem{Alarcon:2012nr}
J.~M. Alarcon, L.~S. Geng, J.~Martin~Camalich and J.~A. Oller, \emph{{The
  strangeness content of the nucleon from effective field theory and
  phenomenology}},
  \href{http://dx.doi.org/10.1016/j.physletb.2014.01.065}{\emph{Phys. Lett.}
  {\bf B730} (2014) 342--346}, [\href{http://arxiv.org/abs/1209.2870}{{\tt
  1209.2870}}].

\bibitem{Akerib:2016vxi}
{\scshape LUX} collaboration, D.~S. Akerib et~al., \emph{{Results from a search
  for dark matter in the complete LUX exposure}},
  \href{http://dx.doi.org/10.1103/PhysRevLett.118.021303}{\emph{Phys. Rev.
  Lett.} {\bf 118} (2017) 021303}, [\href{http://arxiv.org/abs/1608.07648}{{\tt
  1608.07648}}].

\bibitem{Li:2014wia}
T.~Li and Y.-F. Zhou, \emph{{Strongly first order phase transition in the
  singlet fermionic dark matter model after LUX}},
  \href{http://dx.doi.org/10.1007/JHEP07(2014)006}{\emph{JHEP} {\bf 07} (2014)
  006}, [\href{http://arxiv.org/abs/1402.3087}{{\tt 1402.3087}}].

\bibitem{Chao:2017vrq}
W.~Chao, H.-K. Guo and J.~Shu, \emph{{Gravitational Wave Signals of Electroweak
  Phase Transition Triggered by Dark Matter}},
  \href{http://arxiv.org/abs/1702.02698}{{\tt 1702.02698}}.

\bibitem{Cooke:2014}
R.~Cooke, M.~Pettini, R.~A. Jorgenson, M.~T. Murphy and C.~C. Steidel,
  \emph{{Precision measures of the primordial abundance of deuterium}},
  \href{http://dx.doi.org/10.1088/0004-637X/781/1/31}{\emph{Astrophys. J.} {\bf
  781} (2014) 31}, [\href{http://arxiv.org/abs/1308.3240}{{\tt 1308.3240}}].

\bibitem{Agashe:2014kda}
{\scshape Particle Data Group} collaboration, K.~A. Olive et~al., \emph{{Review
  of Particle Physics}},
  \href{http://dx.doi.org/10.1088/1674-1137/38/9/090001}{\emph{Chin. Phys.}
  {\bf C38} (2014) 090001}.

\bibitem{Simha:2008zj}
V.~Simha and G.~Steigman, \emph{{Constraining The Early-Universe Baryon Density
  And Expansion Rate}},
  \href{http://dx.doi.org/10.1088/1475-7516/2008/06/016}{\emph{JCAP} {\bf 0806}
  (2008) 016}, [\href{http://arxiv.org/abs/0803.3465}{{\tt 0803.3465}}].

\bibitem{Delaunay:2007wb}
C.~Delaunay, C.~Grojean and J.~D. Wells, \emph{{Dynamics of Non-renormalizable
  Electroweak Symmetry Breaking}},
  \href{http://dx.doi.org/10.1088/1126-6708/2008/04/029}{\emph{JHEP} {\bf 04}
  (2008) 029}, [\href{http://arxiv.org/abs/0711.2511}{{\tt 0711.2511}}].

\bibitem{Carrington:1991hz}
M.~E. Carrington, \emph{{The Effective potential at finite temperature in the
  Standard Model}},
  \href{http://dx.doi.org/10.1103/PhysRevD.45.2933}{\emph{Phys. Rev.} {\bf D45}
  (1992) 2933--2944}.

\end{thebibliography}\endgroup
%\bibliographystyle{JHEP}
%%%%%%%%%%%%%%%%%%%%%%%%%%%%%%%%%%%%%

\end{document}